\documentclass[floatfix,twocolumn,prl,amsmath]{revtex4}
\usepackage{graphicx}
\usepackage{bm}
\usepackage{amssymb}
\usepackage{float}
\usepackage{dcolumn}
\usepackage{subfigure}
\usepackage{threeparttable}
\usepackage{multirow}
\usepackage{mathrsfs}
\usepackage{simplewick}
\DeclareMathAlphabet{\mathscrbf}{OMS}{mdugm}{b}{n}
\usepackage[usenames,dvipsnames]{color}
\usepackage{mathtools}
\newcommand{\intkspa}{\int\!\!\frac{\rmd^2 k}{(2\pi)^2}}

\begin{document}
\newcommand{\magdir}{\hat{\vn{M}}}
\newcommand{\vn}[1]{{\boldsymbol{#1}}}
\newcommand{\polarivec}{\boldsymbol{\epsilon}}
\newcommand{\crea}[1]{{c_{#1}^{\dagger}}}
\newcommand{\annihi}[1]{{c_{#1}^{\phantom{\dagger}}}}
\newcommand{\vht}[1]{{\boldsymbol{#1}}}
\newcommand{\matn}[1]{{\bf{#1}}}
\newcommand{\matnht}[1]{{\boldsymbol{#1}}}
\newcommand{\bege}{\begin{equation}}
\newcommand{\ee}{\end{equation}}
\newcommand{\bal}{\begin{aligned}}
\newcommand{\defbar}{\overline}
\newcommand{\SM}{\scriptstyle}
\newcommand{\eal}{\end{aligned}}
\newcommand{\torkance}{t}
\newcommand{\rmd}{{\rm d}}
\newcommand{\rme}{{\rm e}}
\newcommand{\udot}{\overset{.}{u}}
\newcommand{\exponential}[1]{{\exp(#1)}}
\newcommand{\phandot}[1]{\overset{\phantom{.}}{#1}}
\newcommand{\phandag}{\phantom{\dagger}}
\newcommand{\Trace}{\text{Tr}}
\newcommand{\Bxc}{\Omega}
\setcounter{secnumdepth}{2}
\title{Spin-orbit torques and tunable Dzyaloshinskii-Moriya interaction in Co/Cu/Co trilayers}
\author{Frank Freimuth}
\email[Corresp.~author:~]{f.freimuth@fz-juelich.de}
\author{Stefan Bl\"ugel}
\author{Yuriy Mokrousov}
\affiliation{Peter Gr\"unberg Institut and Institute for Advanced Simulation,
Forschungszentrum J\"ulich and JARA, 52425 J\"ulich, Germany}
\date{\today}
\begin{abstract}
We study 
the spin-orbit torques (SOTs) 
in Co/Cu/Co magnetic trilayers based on first-principles
density-functional theory calculations in the case where
the applied electric
field lies in-plane, i.e., parallel to the interfaces. We assume that the
bottom Co layer has a fixed in-plane magnetization, while the
top Co layer can be switched. We find that the SOT on the top
ferromagnet can be controlled by the bottom ferromagnet
because of the nonlocal character of the SOT in this system.
As a consequence the SOT is anisotropic, i.e., its magnitude varies
with the
direction of the applied electric field.
We show that the Dzyaloshinskii-Moriya interaction (DMI) in the top layer 
is anisotropic as well, i.e., the spin-spiral wavelength of spin-spirals in the
top layer depends on their in-plane propagation direction. 
This effect suggests that DMI can be tuned easily in magnetic trilayers
via the magnetization direction of the bottom layer.
In order to understand the influence of the 
bottom ferromagnet on the SOTs and the DMI of the top ferromagnet
we study these effects in Co/Cu magnetic bilayers for comparison.
We find the SOTs and the DMI to be surprisingly large despite the
small spin-orbit interaction of Cu.
\end{abstract}

\maketitle
\renewcommand{\arraystretch}{1.3}
\section{Introduction}
In noncentrosymmetric ferromagnets the application of
an electric current generates torques on the magnetization
that result from the spin-orbit interaction (SOI) and that
can be used to switch the 
magnet (see Ref.~\cite{sot_review} for  a recent review).
In magnetic bilayers 
such as Co/Pt, CoFeB/Ta and 
CoFeB/W
with structural inversion asymmetry
spin currents mediate an important contribution to these
so-called
spin-orbit torques (SOTs)~\cite{spin_torque_giant_she_tantalum,stt_devices_giant_she_tungsten,ibcsoit,symmetry_spin_orbit_torques}. 
At first, these spin currents were generated by the
spin Hall effect (SHE) of nonmagnetic heavy metals such as Pt, W, Ta and Pd.
Later, also the antiferromagnets MnPt, MnPd and MnIr were found to be efficient sources of
spin currents
due to their large SHE
angles~\cite{she_in_afms,switching_sot_AFM_FM_bilayer,current_induced_torques_ultrathin_IrMn}. 
Additionally, it has been pointed out
that also the anomalous Hall effect (AHE) and the anisotropic
magnetoresistance (AMR)
of ferromagnets can be used to generate transverse spin currents, because the
transverse charge currents from the AHE and from the planar Hall effect are
spin-polarized~\cite{stt_from_ahe_and_amr}.
Another contribution to the SOT in magnetic bilayers
stems from the interfacial 
SOI~\cite{quantum_kinetic_rashba_macdonald,quantum_kinetic_rashba_manchon,CoPt_Haney_Stiles}.

\begin{figure}
\includegraphics[width=\linewidth,trim=4cm 3cm 4cm 0cm,clip]{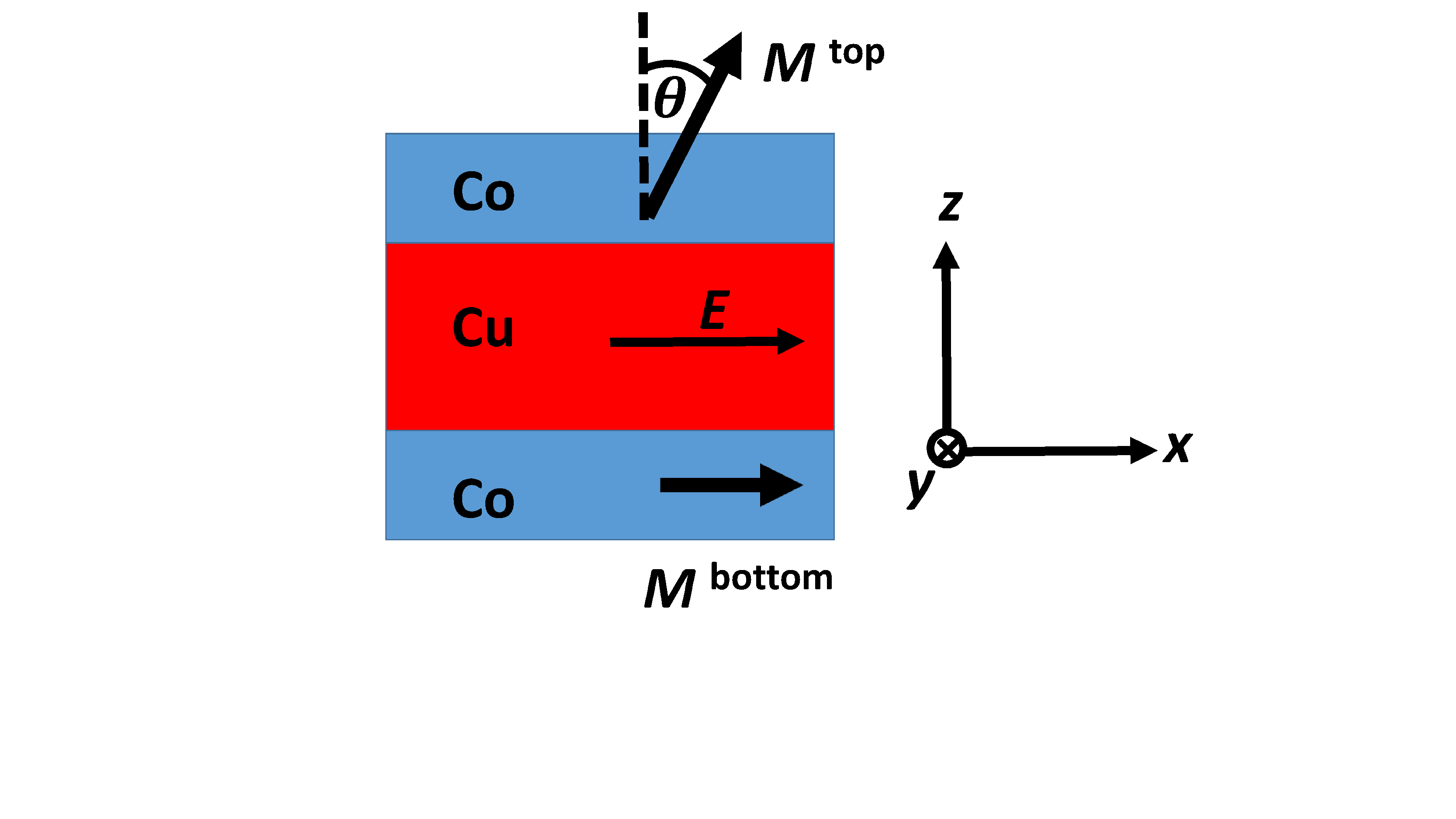}
\caption{\label{cocucotrilayer_setup}
Illustration of a Co/Cu/Co trilayer structure. 
The magnetization  
of the top layer, $\vn{M}^{\rm top}$,
can be switched, while the magnetization of the 
bottom layer, $\vn{M}^{\rm bottom}$, is fixed to point into the $x$ direction.
The two Co-ferromagnets are separated by a Cu normal metal spacer.
The applied electric field $\vn{E}$ points in the in-plane direction, 
e.g.\ in $x$ direction as shown in the figure. The interface normal direction, i.e., the
out-of-plane direction, points in $z$ direction.
}
\end{figure}

Recently, switching of the magnetization by SOT has been demonstrated
in CoFeB/Ti/CoFeB magnetic trilayers~\cite{sots_interface_generated_spin_currents_baek}. 
In the trilayer two CoFeB
ferromagnets (one at the top and one at the bottom) 
are separated by a Ti normal metal 
spacer (see Fig.~\ref{cocucotrilayer_setup} for illustration of
a Co/Cu/Co trilayer). The magnetization of the
bottom ferromagnet (FM) is fixed to the in-plane $x$ direction.
In order to switch the top FM
an electric current is applied parallel to the magnetization
of the bottom FM.
Experiments show that the spin current that switches the top magnet
is generated at the interface between the bottom magnet and the 
normal metal (NM)
spacer and that the spin polarization of the spin current has
components along both the $y$ and the $z$ 
direction~\cite{sots_interface_generated_spin_currents_baek,
observation_spin_orbit_effects_spin_rotation_symmetry}.

Such interface-generated spin currents can be explained in 
terms of spin-orbit filtering and
spin-orbit
precession~\cite{spin_transport_interfaces_soc_formalism,spin_transport_interfaces_soc_phenomenology,interface_generated_spin_currents}
and provide a third mechanism for SOTs, which adds to the
two mechanisms found in magnetic bilayers, i.e., the 
contribution from the SHE and the contribution from the
interfacial SOI at the interface between the top magnet and the
normal metal.
Since the spin polarization of the spin current 
has a component along the $z$ direction,
the magnetization can be switched deterministically
by the applied electric current without the need 
for additional external magnetic fields.
While it has been 
observed in \textit{ab-initio} 
calculations~\cite{invsot,giant_interface_spin_hall} on 
magnetic bilayers that
SHE angles are position-dependent and can
be strongly enhanced close to interfaces, 
the interface-generated spin currents from the
bottom FM interface exhibit a distinct spin polarization,
which can be controlled by the bottom FM magnetization
direction.

Magnetic bilayer structures such as Co/Pt 
are often rotationally symmetric around
the interface normal direction
when the magnetization 
points out-of-plane~\cite{symmetry_spin_orbit_torques}. In such cases
the magnitude of the SOT 
does not depend on the in-plane direction of the applied electric
current. 
However, when the bottom FM magnetization
in a trilayer lies in-plane and the top FM magnetization
points out-of-plane, this symmetry is broken, i.e.,
the SOTs exerted on the top FM are expected to depend on the
direction of the applied electric current.
Since it has been found that SOTs and the
Dzyaloshinskii-Moriya
interaction (DMI) are correlated in various 
ways~\cite{sot_dmi_stiles,common_origin_sot_dmi,mothedmisot},
one may expect that also the DMI is anisotropic in this case
such that the wavelength of spin spirals that are stabilized by DMI depends
on the in-plane direction of the spin-spiral wave vector. 
Equivalently, when a spin spiral in 
the top FM is oriented in a fixed direction, the spin-spiral
wavelength can be tuned by the
magnetization direction in the bottom FM.
Compared to other methods of tuning the DMI, 
which require to generate a strong
non-equilibrium situation by applying 
femtosecond laser 
pulses~\cite{ultrafast_modification_exchange_interaction,spicudmi}, 
the option to tune DMI
via the magnetization direction of the bottom FM
is attractive by its simplicity.

In this work we investigate SOTs and DMI in Co/Cu/Co trilayers
and in Co/Cu bilayers.
These systems are ideal to investigate SOTs that arise from
interface-generated spin currents, because the SHE angle of
bulk Cu is negligibly small and therefore the possibility that there
might be contributions to the SOT originating from the SHE
of bulk Cu can be ruled out.
So far, current-induced torques in Co/Cu/Co spin-valves have been
investigated experimentally~\cite{stt-cocuco_Katine_Albert_Buhrman_Myers_Ralph}
and theoretically~\cite{stt-cocuco_haney&waldron&duine&nunez&guo&macdonald,stt-cocuco_Heiliger_Czerner_Yavorsky_Mertig_Stiles}
only in the current perpendicular to the plane (CPP) geometry.
Only in Co/Cu bilayers the field-like component of the
SOT has been investigated already 
theoretically for the case of electric current applied
parallel to the interfaces~\cite{PhysRevB.93.224420}. 

This paper is organized as follows.
In Sec.~\ref{sec_symmetry} we discuss the symmetry of 
the SOTs in Co/Cu/Co trilayers.
In Sec.~\ref{sec_formalism} we explain the formalism that
we use in order to calculate SOTs and DMI in these systems.
In Sec.~\ref{sec_sot_in_cocuco} we discuss the
\textit{ab-initio} results for the SOTs in
Co/Cu/Cu trilayers and in Co/Cu bilayers.
In Sec.~\ref{sec_dmi_in_cocuco} we present our result on the
DMI in these systems.
This paper ends with a summary in Sec.~\ref{sec_summary}.



\section{Symmetry properties}
\label{sec_symmetry}
We consider a Co/Cu/Co trilayer as illustrated in 
Fig.~\ref{cocucotrilayer_setup}. We assume that the Co and Cu layers are
stacked along their (001) directions.
In this case, 
the crystal lattice of the trilayer
has c4 rotational symmetry around the $z$ axis and
the planes $xz$ and $yz$ are mirror planes of the crystal lattice.
We assume that the magnetization of the bottom 
FM, $\magdir^{\rm bottom}$,
points
into the $x$ direction, while the magnetization direction of the top
FM is given 
by $\magdir^{\rm top}=(\sin\theta\cos\phi,\sin\theta\sin\phi,\cos\theta)^{\rm T}$.
In order to describe the torque $\vn{T}^{\rm top}$ on the top FM we define the
torkance tensor $\vn{t}^{\rm top}$ such that
\bege
\vn{T}^{\rm top}=\vn{t}^{\rm top}\vn{E}.
\ee

We first consider the case with $\theta=0$ and $\phi=0$
when the electric field is applied in $x$ direction.
Since the torque is perpendicular to the magnetization
we only need to consider the two 
components $T_{x}^{\rm top}$ and $T_{y}^{\rm top}$.
The $xz$ mirror plane flips $\magdir^{\rm top}$, $\magdir^{\rm bottom}$,
and $T_{x}^{\rm top}$ (because they are axial vectors),
but it
leaves $E_x$ (polar vector) and $T^{\rm top}_y$ unchanged.
The c2 rotation around the $z$ axis flips all
in-plane vector components.
Thus, combination of the $zx$ mirror plane and the c2 rotation around the $z$ axis 
leaves $T^{\rm top}_x$ and $\magdir^{\rm bottom}$ unchanged, but flips
$E_x$, $T^{\rm top}_y$, and $\magdir^{\rm top}$. Consequently, $T^{\rm top}_x$ is odd 
in $\magdir^{\rm top}$, i.e., $T^{\rm top}_x$ changes sign 
when $\magdir^{\rm top}$ is flipped.
Additionally, $T^{\rm top}_y$ is even in the
magnetization $\magdir^{\rm top}$, i.e., it does not change sign 
when $\magdir^{\rm top}$ is flipped.
When the electric field is applied in $y$ direction, the combination 
of the $zx$ mirror plane and of the c2 rotation around the $z$ axis preserves $E_y$.
Therefore, $T^{\rm top}_x$ is even in $\magdir^{\rm top}$, 
and $T^{\rm top}_y$ is odd in $\magdir^{\rm top}$ when the electric field is applied 
in $y$ direction. 
Since the c2 rotation around the $z$ axis flips the applied electric 
field, $\magdir^{\rm bottom}$, $T^{\rm top}_x$, and $T^{\rm top}_y$, 
it follows that both $T^{\rm top}_x$ and $T^{\rm top}_y$ are even in $\magdir^{\rm bottom}$.
These properties are summarized in the first row of Table~\ref{tab_symmetry}
and in the first row of Table~\ref{tab_symmetry2}.

Next, we consider the case with $\theta=90^{\circ}$ 
and $\phi=0^{\circ}$. 
The magnetizations of both the top FM
and the bottom FM point in $x$ direction
in this case. 
Since the torque is perpendicular to the magnetization
we only consider $T^{\rm top}_y$ and $T^{\rm top}_z$.
We first assume that the electric field is applied in $x$ direction.
The $xz$ mirror plane preserves $E_x$ and $T_y^{\rm top}$, 
but it flips $T_z^{\rm top}$, $\magdir^{\rm top}$ and $\magdir^{\rm bottom}$.
Consequently, $T_y^{\rm top}$ does not change
when both $\magdir^{\rm top}$ and $\magdir^{\rm bottom}$ are flipped, 
while $T_z^{\rm top}$ changes sign in this case.
When the electric field is applied in $y$ direction,
the $yz$ mirror plane preserves $E_y$, $\magdir^{\rm top}$,
and $\magdir^{\rm bottom}$, but it flips 
both $T^{\rm top}_y$ and $T^{\rm top}_z$. Thus,
symmetry requires $t^{\rm top}_{yy}=0$ and $t^{\rm top}_{zy}=0$.
These properties are summarized in the second row of Table~\ref{tab_symmetry2}.

\begin{threeparttable}
\caption{Symmetry properties of the torkance tensor $t_{ij}^{\rm top}$
for various directions of the 
magnetization $\magdir^{\rm top}=(\sin\theta\cos\phi,\sin\theta\sin\phi,\cos\theta)^{\rm T}$.
$+$ means that the torkance is even in $\magdir^{\rm top}$
(i.e., it does not change sign when $\magdir^{\rm top}$ is flipped 
while $\magdir^{\rm bottom}$ is not flipped) and
$-$ means that the torkance is odd in $\magdir^{\rm top}$ 
(i.e., it changes sign when $\magdir^{\rm top}$ is flipped 
while $\magdir^{\rm bottom}$ is not flipped).
}
\label{tab_symmetry}
\begin{ruledtabular}
\begin{tabular}{c|c|c|c|c|c|c|}
&$t_{xx}^{\rm top}$
&$t_{xy}^{\rm top}$
&$t_{yx}^{\rm top}$
&$t_{yy}^{\rm top}$
&$t_{zx}^{\rm top}$
&$t_{zy}^{\rm top}$
\\
\hline
$\theta=0,\phi=0$ 
&-&+&+&-&0&0\\
\hline
$\theta=90^{\circ},\phi=90^{\circ}$
&-&+&0&0&+&-\\
\end{tabular}
\end{ruledtabular}
\end{threeparttable}

\begin{threeparttable}
\caption{Symmetry properties of the torkance tensor $t_{ij}^{\rm top}$
for various directions of the 
magnetization $\magdir^{\rm top}=(\sin\theta\cos\phi,\sin\theta\sin\phi,\cos\theta)^{\rm T}$.
$+$ means that the torkance does not change sign when 
both $\magdir^{\rm top}$ and $\magdir^{\rm bottom}$ are flipped.
$-$ means that the torkance changes sign when 
both $\magdir^{\rm top}$ and $\magdir^{\rm bottom}$ are flipped.
}
\label{tab_symmetry2}
\begin{ruledtabular}
\begin{tabular}{c|c|c|c|c|c|c|}
&$t_{xx}^{\rm top}$
&$t_{xy}^{\rm top}$
&$t_{yx}^{\rm top}$
&$t_{yy}^{\rm top}$
&$t_{zx}^{\rm top}$
&$t_{zy}^{\rm top}$
\\
\hline
$\theta=0,\phi=0$ 
&-&+&+&-&0&0\\
\hline
$\theta=90^{\circ},\phi=0$
&0&0&+&0&-&0\\
\hline
$\theta=90^{\circ},\phi=90^{\circ}$
&+&+&0&0&-&-\\
\end{tabular}
\end{ruledtabular}
\end{threeparttable}

Finally, we consider the case where $\magdir^{\rm top}$
points in $y$ direction, i.e., $\theta=90^{\circ}$ and $\phi=90^{\circ}$.
Since the torque is perpendicular to the magnetization we only need to
consider $T^{\rm top}_x$ and $T^{\rm top}_z$.
First, we assume that the electric field is applied in $x$ direction.
The $xz$ mirror plane flips $T_x^{\rm top}$, $T_z^{\rm top}$ and $\magdir^{\rm bottom}$,
but it preserves $E_x$ and $\magdir^{\rm top}$. 
Thus, the combination of the c2 rotation around the $z$ axis with 
the $xz$ mirror plane 
preserves $T_x^{\rm top}$ and $\magdir^{\rm bottom}$
but flips $E_x$, $\magdir^{\rm top}$ and $T_z^{\rm top}$.
Consequently, $T_z^{\rm top}$ is even in $\magdir^{\rm top}$
and $T_x^{\rm top}$ is odd in $\magdir^{\rm top}$.
When the electric field points in $y$ direction,
the combination of the c2 rotation around the $z$ axis with 
the $xz$ mirror plane preserves $E_y$.
In this case $T_z^{\rm top}$ is odd in $\magdir^{\rm top}$
and $T_x^{\rm top}$ is even in $\magdir^{\rm top}$.
A c2 rotation around the $z$ axis 
flips $\magdir^{\rm top}$, $\magdir^{\rm bottom}$, $\vn{E}$, 
and $T_x^{\rm top}$ but preserves $T_z^{\rm top}$.
Consequently, $T_x^{\rm top}$ does not change sign when
both $\magdir^{\rm top}$ and $\magdir^{\rm bottom}$ are flipped,
while $T_z^{\rm top}$ changes sign when both $\magdir^{\rm top}$ 
and $\magdir^{\rm bottom}$ are flipped.
These properties are summarized 
in the second row of Table~\ref{tab_symmetry}
and
in the third row of 
Table~\ref{tab_symmetry2}.

The
case $\theta=90^{\circ}$ and $\phi=90^{\circ}$
is of special interest,
because it is well-known for magnetic bilayers
that $t_{zx}=0$ and $t_{xx}=0$ when $\theta=90^{\circ}$ 
and $\phi=90^{\circ}$~\cite{symmetry_spin_orbit_torques}.
Indeed, from Table~\ref{tab_symmetry}
and
Table~\ref{tab_symmetry2}
it is clear that $t_{zx}^{\rm top}$ and $t_{xx}^{\rm top}$
cannot exist without
the bottom magnetic layer: 
For $\theta=90^{\circ}$ and $\phi=90^{\circ}$
$t_{zx}^{\rm top}$ is even in 
the magnetization of the top layer (see Table~\ref{tab_symmetry}), 
but odd in the total
magnetization (see Table~\ref{tab_symmetry2}). 
Similarly, $t_{xx}^{\rm top}$ is odd in 
the magnetization of the top layer (see Table~\ref{tab_symmetry}), 
but even in the total
magnetization (see Table~\ref{tab_symmetry2}). 
Clearly, these conditions cannot be satisfied when the
bottom magnetic layer is missing.
Since $t_{zx}^{\rm top}=t_{xx}^{\rm top}=0$ in bilayers, the underlying mechanism
in trilayers has to involve the bottom FM, i.e.,
it has to be a nonlocal mechanism.
Since nonlocal transfer of angular momentum is mediated
by spin currents, we attribute the nonzero $t_{zx}^{\rm top}$ in trilayers 
to a spin current with
spin-polarization along $z$ direction flowing from the
bottom FM to the top FM. Such a spin current can be
generated at the bottom FM interface through 
spin-orbit precession~\cite{interface_generated_spin_currents,spin_transport_interfaces_soc_formalism,sots_interface_generated_spin_currents_baek}.
$t_{zx}^{\rm top}$ is odd in the magnetization of the bottom magnet,
consistent with the mechanism of spin-orbit precession.
As the $z$-polarized spin current interacts with the top FM it
precesses around the top FM magnetization. This leads to an
additional torque 
component in the direction of $\hat{\vn{M}}^{\rm top}\times \hat{\vn{e}}_{z}$,
which is odd in $\hat{\vn{M}}^{\rm top}$.
For  $\hat{\vn{M}}^{\rm top}=\hat{\vn{e}}_{y}$ this torque points
into the $x$ direction, which explains the component $t_{xx}^{\rm top}$.
Therefore, both $t_{xx}^{\rm top}$ and $t_{zx}^{\rm top}$ arise from 
the $z$-polarized spin current. Since $t_{xx}^{\rm top}$ is odd in 
the top FM magnetization,
it cannot be explained by a spin current with spin-polarization along $x$
that interacts with the top FM through spin transfer, because the
resulting torque would be even in $\hat{\vn{M}}^{\rm top}$.
The $z$-polarized spin current in trilayers is of particular
interest, because it allows field-free 
switching and provides an antidamping-torque for the
perpendicular top 
FM~\cite{sots_interface_generated_spin_currents_baek,observation_spin_orbit_effects_spin_rotation_symmetry}.
Calculating the SOT at $\theta=\phi=90^{\circ}$ is
an easy way to investigate this $z$-polarized
spin current. In comparison, the SOT at $\phi=0^{\circ}$
and $0<\theta<90^{\circ}$ contains also contributions from the
interfacial SOI at the top FM interface and therefore it does not
give direct access to the $z$-polarized spin current. 

\section{Formalism}
\label{sec_formalism}
\subsection{Spin-orbit torque}
We define the torkance $t_{ij\alpha}$ of atom $\alpha$ by
the
equation 
\bege\label{eq_torque_and_torkance}
T_{i\alpha}=\sum_{j}
t_{ij\alpha}E_j,
\ee
where $E_j$ is the $j$-th component of the applied electric field
and $T_{i\alpha}$ is the $i$-th component of the torque exerted
on the magnetic moment of atom $\alpha$. 
In the Co/Cu/Co trilayers considered in this
work we will be particularly interested in the sum 
of $t_{ij\alpha}$
over all atoms in the
top FM, 
\bege\label{eq_sot_toplayer}
t_{ij}^{\rm top}=
\sum_{\alpha\in {\rm top}}
t_{ij\alpha},
\ee
where $\alpha\in {\rm top}$ denotes the atoms in the top FM.
Since the bottom and top FM are separated by a Cu spacer,
the top FM can switch while the bottom FM can be
kept fixed. For the magnetization dynamics of the top FM
the sum as defined in Eq.~\eqref{eq_sot_toplayer} is the relevant torkance.
In magnetic bilayers with only one ferromagnetic layer,
such as a Co/Cu magnetic bilayer, the relevant
torkance is obtained by summing over all atoms, i.e.,
\bege\label{eq_sot_total}
t_{ij}=
\sum_{\alpha}
t_{ij\alpha}.
\ee

The torkance is given by the sum
of three terms, i.e.,
$t^{\phantom{II}}_{ij\alpha}=t^{\rm I(a)}_{ij\alpha}+
t^{\rm I(b)}_{ij\alpha}+
t^{\rm II}_{ij\alpha}
$,
where~\cite{ibcsoit,invsot}
\begin{gather}\label{eq_torkance_kubo}
\begin{aligned}
t^{\rm I(a)\phantom{I}}_{ij\alpha}\!\!\!\!&=\phantom{-}\frac{eA}{h}
\intkspa
\,{\rm Tr}
\left\langle
\mathcal{T}_{i\alpha}
G^{\rm R}_{\vn{k}}(\mathcal{E}_{\rm F})
v_{j}
G^{\rm A}_{\vn{k}}(\mathcal{E}_{\rm F})
\right\rangle
\\
t^{\rm I(b)\phantom{I}}_{ij\alpha}\!\!\!\!&=-\frac{eA}{h}
\intkspa
\,{\rm Re}
\,{\rm Tr}
\left\langle
\mathcal{T}_{i\alpha}
G^{\rm R}_{\vn{k}}(\mathcal{E}_{\rm F})
v_{j}
G^{\rm R}_{\vn{k}}(\mathcal{E}_{\rm F})
\right\rangle
\\
t^{\rm II\phantom{(a)}}_{ij\alpha}\!\!\!\!&=
\phantom{-}\frac{eA}{h}\int_{-\infty}^{\mathcal{E}_{\rm F}}
d\mathcal{E}
\intkspa
\,{\rm Re}
\,{\rm Tr}
\left\langle
\mathcal{T}_{i\alpha}G_{\vn{k}}^{\rm R}(\mathcal{E})v_{j}
\frac{dG_{\vn{k}}^{\rm R}(\mathcal{E})}{d\mathcal{E}}\right.\\
 &\quad\quad\quad\quad\quad\quad\quad\quad\,-\left.
\mathcal{T}_{i\alpha}\frac{dG_{\vn{k}}^{\rm R}(\mathcal{E})}{d\mathcal{E}}v_{j}G_{\vn{k}}^{\rm R}(\mathcal{E})
\right\rangle.
\end{aligned}
\end{gather}
Here, $A$ is the area of
the in-plane unit cell, such that $T_{i\alpha}$
in Eq.~\eqref{eq_torque_and_torkance} is the $i$-th component
of the torque exerted on the magnetic moments of atom type $\alpha$
in one unit cell.
$e>0$ is the elementary positive charge, $\vn{k}$ denotes a $k$-point in
the two-dimensional Brillouin zone of the bilayers and trilayers,
 $G_{\vn{k}}^{\rm R}(\mathcal{E})$ is the
retarded Green's function,
$G^{\rm A}_{\vn{k}}(\mathcal{E})$ is the advanced Green's 
function, $\mathcal{E}_{\rm F}$ is the Fermi energy,
$v_{j}$ is the $j$-th component of the velocity operator, and
$\mathcal{T}_{i\alpha}$ is the $i$-th component of the
torque operator $\vn{\mathcal{T}}_{\!\!\alpha}$ of 
atom $\alpha$
with matrix elements 
\bege\label{eq_atom_resolved_torque_operator}
\langle\psi^{\phantom{\dagger}}_{\vn{k}n}
|
\vht{\mathcal{T}}^{\phantom{\dagger}}_{\!\!\alpha}
|
\psi^{\phantom{\dagger}}_{\vn{k}m}\rangle
=-\mu^{\phantom{\dagger}}_{\rm B}\!\!\!\!\int\limits_{\rm MT_{\alpha}}\!\!\!\!d^3r\,
\psi^{\dagger}_{\vn{k}n}(\vn{r})
\vht{\sigma}\!\times\! \vn{\Bxc}^{\rm xc}(\vn{r})
\psi^{\phantom{\dagger}}_{\vn{k}m}(\vn{r}),
\ee
where $\mu_{\rm B}$ is the Bohr magneton, $\vn{\sigma}$
is the vector of Pauli spin matrices,
$\vn{\Bxc}^{\rm xc}(\vn{r})$ is the exchange field, which 
is obtained from the difference between the effective potentials
of minority and majority electrons 
as $\vn{\Bxc}^{\rm xc}(\vn{r})=\frac{1}{2\mu_{\rm B}}\left[
V^{\rm eff}_{\rm minority}(\vn{r})-V^{\rm eff}_{\rm majority}(\vn{r})
\right]\magdir(\vn{r})$, and $\magdir(\vn{r})$ is the magnetization direction. 
The volume
integration in Eq.~\eqref{eq_atom_resolved_torque_operator}
is restricted to the MT-sphere of atom $\alpha$.

The torkances depend on the magnetization direction $\magdir$.
We define the
even torkance $t^{\rm e}_{ij\alpha}(\magdir)=[t_{ij\alpha}(\magdir)+t_{ij\alpha}(-\magdir)]/2$
and the odd torkance $t^{\rm o}_{ij\alpha}(\magdir)=[t_{ij\alpha}(\magdir)-t_{ij\alpha}(-\magdir)]/2$.
Inversion of 
magnetization, i.e., $\magdir\rightarrow -\magdir$,
does not modify the even torkance, 
but it flips the sign of the odd torkance.
The torkance is the sum of its even and odd parts, 
i.e., $t^{\phantom{e}}_{ij\alpha}(\magdir)=t^{\rm e}_{ij\alpha}(\magdir)+t^{\rm o}_{ij\alpha}(\magdir)$.
$t^{\rm II\phantom{(a)}}_{ij\alpha}$ in Eq.~\eqref{eq_torkance_kubo}
contributes only to $t^{\rm e}_{ij\alpha}$ 
and $t^{\rm I(b)\phantom{I}}_{ij\alpha}$ contributes only to $t^{\rm o}_{ij\alpha}$, 
while $t^{\rm I(a)\phantom{I}}_{ij\alpha}$ contributes to 
both $t^{\rm e}_{ij\alpha}$ 
and $t^{\rm o}_{ij\alpha}$.
Since the magnetizations in the top and bottom FMs may point into different
directions in trilayers (see Fig.~\ref{cocucotrilayer_setup}), it is important to point out
that $t^{\rm o}_{ij\alpha}(\magdir)$ changes sign when both $\magdir^{\rm top}$
and $\magdir^{\rm bottom}$ are flipped, 
while $t^{\rm e}_{ij\alpha}(\magdir)$ does not change sign when both $\magdir^{\rm top}$
and $\magdir^{\rm bottom}$ are flipped, because the odd character of $t^{\rm o}_{ij\alpha}(\magdir)$ 
and the even character of $t^{\rm e}_{ij\alpha}(\magdir)$ result from their transformation
properties under time reversal~\cite{ibcsoit}.
Therefore, 
the even torkance in Co/Cu/Co trilayers corresponds to the
plus entries in Table~\ref{tab_symmetry2} and the 
odd torkance corresponds to the minus entries in Table~\ref{tab_symmetry2}.
As discussed in Table~\ref{tab_symmetry} one can also consider how the torkance
changes when only $\magdir^{\rm top}$ is flipped while $\magdir^{\rm bottom}$
is kept fixed and in some cases the torkance for $-\magdir^{\rm top}$ and $+\magdir^{\rm bottom}$ 
is simply equal or opposite to the torkance for $+\magdir^{\rm top}$ and $+\magdir^{\rm bottom}$.
Thus, one can also define even and odd torques with respect to $\magdir^{\rm top}$ when
$\magdir^{\rm bottom}$ is kept fixed. The fact that the third row in 
Table~\ref{tab_symmetry2} differs from the second row in Table~\ref{tab_symmetry} shows
that flipping the bottom magnetization $\magdir^{\rm bottom}$ may impact the torque on the top magnet.
Therefore, in the following we mean by even torque always that the torque does not change when 
both $\magdir^{\rm top}$ and $\magdir^{\rm bottom}$ are flipped, while by odd torque we mean
that the torque changes sign when both $\magdir^{\rm top}$ and $\magdir^{\rm bottom}$ are flipped.

\subsection{Dzyaloshinskii-Moriya interaction}
DMI is a chiral interaction, which modifies the free energy of 
noncentrosymmetric magnets proportional to
the spatial derivatives of the magnetization.
The contribution of DMI to the free energy density 
at position $\vn{r}$ can be
written as~\cite{mothedmisot}
\bege\label{eq_free_energy_dmi}
F^{\rm DMI}(\vn{r})=\sum_{ij}
D_{ij}(\magdir(\vn{r}))
\hat{\vn{e}}_{i}\cdot
\left[
\magdir(\vn{r})
\times\frac{\partial 
\magdir(\vn{r})
}{\partial r_j}
\right],
\ee
where $D_{ij}(\magdir)$ are the DMI coefficients and $\hat{\vn{e}}_{i}$ is a
unit vector pointing in the $i$th spatial direction.
These DMI coefficients can be decomposed into the contributions
associated with the magnetic moments of atom type $\alpha$,
which we denote by $D_{ij\alpha}$,
such that
\bege
D_{ij}=\sum_{\alpha}D_{ij\alpha}.
\ee
In magnetic trilayers such as Co/Cu/Co we will be 
particularly interested in summing $D_{ij\alpha}$ over the layers
of the top magnet in order to quantify the free energy gained 
due to the DMI
when the magnetization in the top layer forms a cycloidal spin spiral
while the magnetization in the bottom layer remains unchanged,
i.e., we will be interested in the quantity
\bege\label{eq_dmi_toplayer}
D^{\rm top}_{ij}=\sum_{\alpha\in{\rm top}}D_{ij\alpha}.
\ee  
We evaluate the DMI coefficients associated with the magnetic moments 
of atom type $\alpha$ from~\cite{mothedmisot,phase_space_berry,itsot}
\bege\label{eq_dmi}
D_{ij\alpha}=\intkspa
\!\!\sum_{n}
f_{\vn{k}n}
\left[
A_{\vn{k}nij\alpha}-(\mathcal{E}_{\vn{k}n}-\mu)B_{\vn{k}nij\alpha}
\right],
\ee
where
\bege\label{eq_akn_kubo}
A_{\vn{k}nij\alpha}=\hbar\sum_{m\neq n}\text{Im}
\left[
\frac{
\langle \psi_{\vn{k}n}  |\mathcal{T}_{i\alpha}| \psi_{\vn{k}m}  \rangle
\langle \psi_{\vn{k}m}  |v_{j}| \psi_{\vn{k}n}  \rangle
}
{
\mathcal{E}_{\vn{k}m}-\mathcal{E}_{\vn{k}n}
}
\right]
\ee
and
\bege\label{eq_bkn_kubo}
B_{\vn{k}nij\alpha}
\!=\!-2\hbar\!\!
\sum_{m\neq n}\!\!\text{Im}
\!\!\left[\!
\frac{
\langle \psi_{\vn{k}n}  |\mathcal{T}_{i\alpha}| \psi_{\vn{k}m}  \rangle
\langle \psi_{\vn{k}m}  |v_{j}| \psi_{\vn{k}n}  \rangle
}
{
(\mathcal{E}_{\vn{k}m}-\mathcal{E}_{\vn{k}n})^2
}
\!\right].
\ee
Here, $\mathcal{E}_{\vn{k}n}$ is the band energy of state $| \psi_{\vn{k}n}  \rangle$.
Eq.~\eqref{eq_dmi} is valid at zero temperature, which is sufficient for this study.
The generalization to finite temperatures has been described 
elsewhere~\cite{mothedmisot,phase_space_berry,itsot}.
Eq.~\eqref{eq_dmi}, Eq.~\eqref{eq_akn_kubo},
and Eq.~\eqref{eq_bkn_kubo} differ from previous 
works~\cite{mothedmisot,phase_space_berry,itsot} 
by the introduction of the atomic index $\alpha$ and by the use of the
atom-resolved torque operator $\mathcal{T}_{i\alpha}$ defined in 
Eq.~\eqref{eq_atom_resolved_torque_operator}.
This generalization is necessary in order to study the DMI associated
with one magnetic layer, e.g.\ for the top magnet according to Eq.~\eqref{eq_dmi_toplayer}, 
in a magnetic multilayer structure.
\section{SOT in C\lowercase{o}/C\lowercase{u}/C\lowercase{o} trilayers}
\label{sec_sot_in_cocuco}
\subsection{Computational details}
We calculate SOTs in magnetic trilayers composed of 3 atomic layers
of Co, 9 atomic layers of Cu, and 3 atomic layers of Co (abbreviated as
Co(3)/Cu(9)/Co(3) in the following). Additionally we compute trilayers
with only 3 or 6 instead of 9 Cu layers, which we abbreviate in the following as
Co(3)/Cu(3)/Co(3) or Co(3)/Cu(6)/Co(3), respectively.
In order to be able to interpret the results on the trilayers
we also calculate a Co/Cu bilayer for
comparison, which is
composed of 3 layers of
Co on 9 layers of Cu (abbreviated as Co(3)/Cu(9)).
In order to calculate the SOTs we first perform
an electronic structure calculation with the FLEUR~\cite{fleurcode} code
using the generalized gradient approximation~\cite{PerdewBurkeEnzerhof}.
We choose the computational unit cell similar to the one in
Ref.~\cite{stt-cocuco_haney&waldron&duine&nunez&guo&macdonald},
where the spin-transfer torque in the current-perpendicular-to-plane-geometry 
was investigated.
However,  
we do not embed the scattering region between semi-infinite
leads as in Ref.~\cite{stt-cocuco_haney&waldron&duine&nunez&guo&macdonald},
but instead we set up the system in the film geometry, i.e., we calculate
Vacuum/Co(3)/Cu(n)/Co(3)/Vacuum (n=3,6,9), 
because we study the SOTs driven by an electric
current that is applied parallel to the interfaces.
For this purpose we use the
film mode of FLEUR, which explicitly takes the vacuum regions into
account~\cite{Krakauer_film_mode}.
We choose the coordinate system such that the $z$ axis is perpendicular to
the interfaces, while the $x$ and $y$ axes are parallel to the 
interfaces (see also Fig.~\ref{cocucotrilayer_setup}). 
Since the magnetization
in the top FM needs to point into a different direction than
the magnetization in the bottom FM of the Co/Cu/Co trilayer,
we use the noncollinear mode of FLEUR~\cite{noco_flapw}.
The magnetization of the bottom FM points in $x$ direction
in our calculations. We specify the magnetization direction
of the top FM in terms of angles $\theta$ and $\phi$, 
i.e., $\vn{m}=(\sin\theta\cos\phi,\sin\theta\sin\phi,\cos\theta)^{\rm T}$.
We include SOI in the calculations.
The computational unit cell of the
Co(3)/Cu(9)/Co(3) trilayer contains 15 atoms in total.
Both Co and Cu are treated as fcc with lattice 
constant 3.54~\AA. Consequently, the
distance between adjacent planes is 1.77~\AA.
The MT-radii of Co and Cu are set to 1.22~\AA.

We label the three Co-layers in the top FM as
Co-1, Co-2, and Co-3, where Co-1 is adjacent to vacuum and
Co-3 faces the Cu-layer.
We denote the three Co-layers in the bottom FM
by Co-4, Co-5, and Co-6, where Co-4 is adjacent to the Cu-layer
and Co-6 faces vacuum.
In the Co(3)/Cu(9)/Co(3) trilayer
the Cu-layers are labelled Cu-1 through Cu-9, where Cu-1 is adjacent
to Co-3 of the top FM.

As explained in Ref.~\cite{ibcsoit}
we use Wannier interpolation~\cite{rmp_wannier90} 
in order to
evaluate Eq.~\eqref{eq_torkance_kubo} for the SOT
and Eq.~\eqref{eq_dmi} for the DMI computationally efficiently.
For this purpose, we generate
18 maximally localized Wannier functions (MLWFs) per atom
using our interface~\cite{WannierPaper} 
between the FLEUR code and 
the Wannier90 code~\cite{wannier90}.
For the generation of MLWFs we used an $8\times 8$ $\vn{k}$ mesh and
as trial orbitals we chose 6 $sp^3d^2$, $d_{xy}$, $d_{yz}$, and $d_{zx}$ for 
spin-up and for spin-down, i.e., 18 trial orbitals per atom.
In the case of the Co(3)/Cu(9)/Co(3) trilayer we generated 270 MLWFs,
which we disentangled from 378 bands.
In the case of the Co(3)/Cu(9) bilayer we generated 216 MLWFs, which we disentangled
from 304 bands.

In order to investigate the role of spin-orbit coupling in the normal metal spacer
layer we also compute the SOTs in Co(3)/Pt(13)/Co(3) trilayers, where Pt
provides strong SOI. The computational details of the Co/Pt/Co trilayer are
given in Ref.~\cite{ibcsoit}.

\subsection{Even SOT}
\begin{figure}
\includegraphics[width=\linewidth]{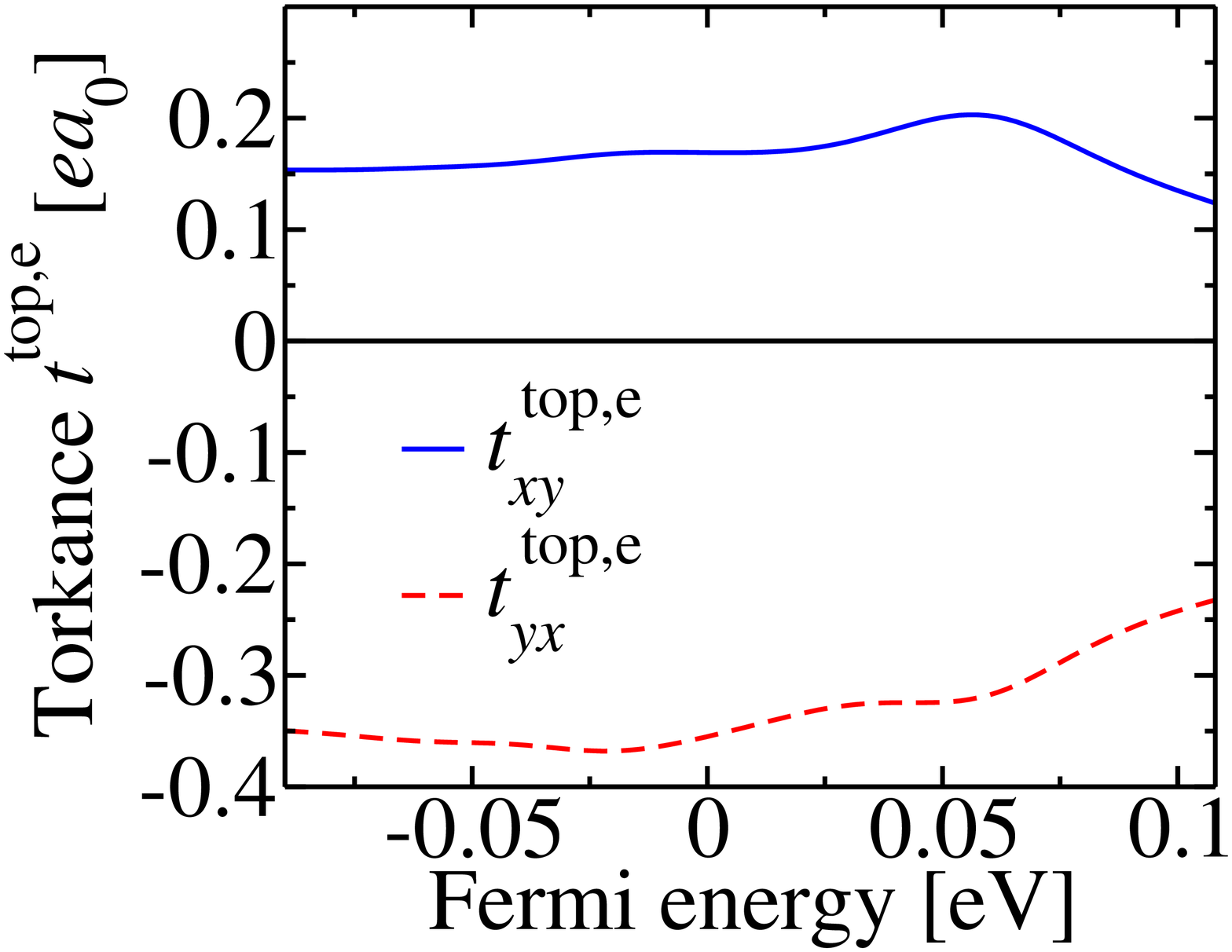}
\caption{\label{cocuco_eventorque_vs_fermi}
Even torkance $t^{\rm top,e}_{ij}$ on 
the top magnetic layer in the
Co(3)/Cu(9)/Co(3) magnetic trilayer 
vs.\ Fermi energy $\mathcal{E}_{\rm F}$
when $\theta=\phi=0$ and $\Gamma=25$meV.
}
\end{figure}

In Fig.~\ref{cocuco_eventorque_vs_fermi} we show the
even torkance of the top FM (see Eq.~\eqref{eq_sot_toplayer}) 
in the Co(3)/Cu(9)/Co(3) trilayer as a function of
the Fermi energy $\mathcal{E}_{\rm F}$ for a lifetime broadening 
of $\Gamma=$25meV when the magnetization of the top FM
points in $z$ direction, i.e., $\theta=\phi=0$. 
By plotting the results as a function of $\mathcal{E}_{\rm F}$ we
can roughly anticipate chemical trends: Negative $\mathcal{E}_{\rm F}$
mimics the system Fe$_{x}$Co$_{1-x}$/Ni$_{x}$Cu$_{1-x}$/Fe$_{x}$Co$_{1-x}$
while positive $\mathcal{E}_{\rm F}$ mimics the system
Ni$_{x}$Co$_{1-x}$/Zn$_{x}$Cu$_{1-x}$/Ni$_{x}$Co$_{1-x}$.
The pure Co(3)/Cu(9)/Co(3) system corresponds to $\mathcal{E}_{\rm F}=0$.
In agreement with the symmetry analysis in Table~\ref{tab_symmetry2}
the two nonzero components of the even torkance are $t_{xy}^{\rm top,e}$
and $t_{yx}^{\rm top,e}$.  
Due to the presence of the bottom FM with magnetization
in $x$ direction the $x$ and $y$ directions are not equivalent and therefore
the torkance components $t^{\rm top,e}_{xy}$ and $t^{\rm top,e}_{yx}$ differ 
substantially (in case of equivalent $x$ 
and $y$ directions $t_{xy}=-t_{yx}$ would hold).
For the range of $\mathcal{E}_{\rm F}$ shown in the 
Figure, $t^{\rm top,e}_{yx}$ is much larger than $t^{\rm top,e}_{xy}$.
This means that when the electric field is applied in $x$ direction, i.e., parallel to
the magnetization of the bottom FM, 
the SOT is much larger than when the electric field is applied perpendicular to
the magnetization of the bottom FM.

\begin{figure}
\includegraphics[width=\linewidth]{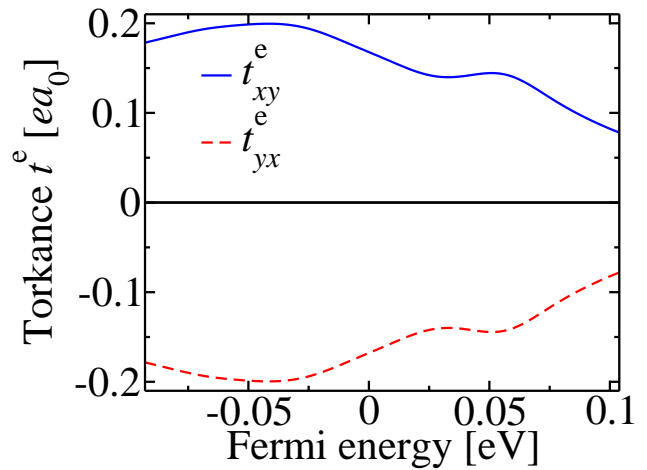}
\caption{\label{eventorque_cocubilayer_vs_fermi}
Even torkance $t^{\rm e}_{ij}$ 
in the Co(3)/Cu(9) magnetic bilayer
vs.\ Fermi energy $\mathcal{E}_{\rm F}$
when $\theta=\phi=0$ and $\Gamma=25$meV.
}
\end{figure}

For comparison we show 
in Fig.~\ref{eventorque_cocubilayer_vs_fermi} the 
total even torkance (see Eq.~\eqref{eq_sot_total}) 
of the Co(3)/Cu(9) bilayer as a function of
Fermi energy $\mathcal{E}_{\rm F}$ for a lifetime broadening of $\Gamma=$25meV
when the magnetization points in $z$ direction.
In this case, the SOT in the Co(3)/Cu(9) bilayer is 
isotropic, i.e., $t^{\rm e}_{xy}=-t^{\rm e}_{yx}$. 
The component $t^{\rm e}_{xy}$ in Co(3)/Cu(9) is very similar in size 
to $t^{\rm top,e}_{xy}$ in Co(3)/Cu(9)/Co(3), 
but $t^{\rm top,e}_{yx}$ in Co(3)/Cu(9)/Co(3) is 
much larger than $t^{\rm e}_{yx}$ in Co(3)/Cu(9).
Consequently, the bottom magnet
with magnetization along $x$
in the Co(3)/Cu(9)/Co(3) trilayer strongly enhances $t^{\rm top,e}_{yx}$, while its
effect on $t^{\rm top,e}_{xy}$ is not so strong.

\begin{figure}
\includegraphics[width=\linewidth]{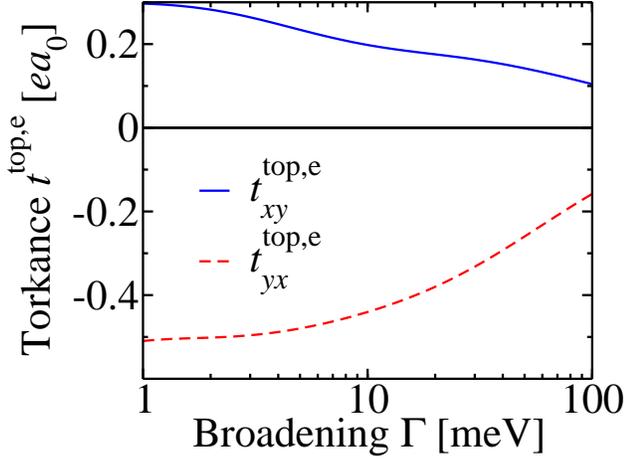}
\caption{\label{eventorque_cocucotrilayer_vs_sigma}
Even torkance $t^{\rm top,e}_{ij}$ on 
the top magnetic layer in the
Co(3)/Cu(9)/Co(3) magnetic trilayer
vs.\ lifetime broadening $\Gamma$
when $\theta=\phi=0$.
}
\end{figure}

In Fig.~\ref{eventorque_cocucotrilayer_vs_sigma} we show the
even torkance of the top FM (see Eq.~\eqref{eq_sot_toplayer}) 
in the Co(3)/Cu(9)/Co(3) trilayer as a function of
the lifetime broadening $\Gamma$ when the magnetization 
points in $z$ direction ($\theta=0$). 
At small broadening $\Gamma=1$~meV the component $t^{\rm top,e}_{yx}$
is larger than $t^{\rm top,e}_{xy}$ by a factor of 1.7 and at
large broadening $\Gamma=100$~meV the component $t^{\rm top,e}_{yx}$
is larger than $t^{\rm top,e}_{xy}$ by a factor of 1.5. Thus, with increasing $\Gamma$
the influence of the bottom FM on $t^{\rm top,e}_{ij}$ becomes smaller
such that the anisotropy of the SOT reduces.
This can be explained by assuming that the effect of the bottom FM
on $t^{\rm top,e}_{ij}$ is mediated by spin currents that flow between the
bottom FM and the top FM. Such a spin-current mediated
effect is not possible when the spin-diffusion length is much smaller than the
thickness of the Cu spacer layer. Since the spin-diffusion length decreases with
increasing disorder, which we model by the broadening $\Gamma$, it is 
expected that an increase of $\Gamma$ reduces the spin-current-mediated
coupling between the bottom and top FMs.
The enhancement of $t^{\rm top,e}_{yx}$ in comparison 
to $t^{\rm top,e}_{xy}$ can be explained by a spin current with spin polarization
in $y$ direction, which flows from the bottom FM into the top FM.
Such a spin current can be generated through spin-orbit filtering
at the bottom FM interface~\cite{interface_generated_spin_currents}.
According to Table~\ref{tab_symmetry} and 
Table~\ref{tab_symmetry2}
$t^{\rm top,e}_{yx}$ is even in $\hat{\vn{M}}^{\rm top}$ and
even in $\hat{\vn{M}}^{\rm bottom}$, consistent with this 
mechanism. In order to describe the spin-orbit
filtering quantitatively, a model has been proposed that predicts this
contribution to be proportional to the relaxation 
time $\tau$~\cite{interface_generated_spin_currents}. 
Since $\tau\propto \hbar/(2\Gamma)$, one expects the contribution
from spin-orbit filtering to decrease with increasing $\Gamma$.
Therefore, there are two reasons why the enhancement 
of $t^{\rm top,e}_{yx}$ in comparison to $t^{\rm top,e}_{xy}$ decreases with
increasing $\Gamma$: The spin-current is attenuated stronger on the 
way from the bottom FM to the top FM and the amount of spin current
generated at the bottom FM interface through spin-orbit filtering decreases.

\begin{figure}
\includegraphics[width=\linewidth]{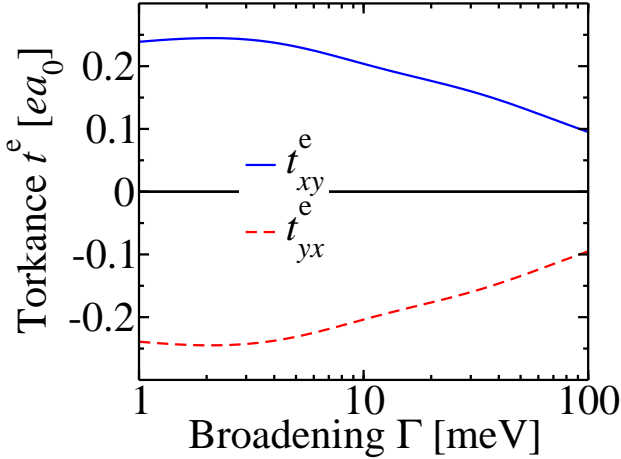}
\caption{\label{eventorque_cocubilayer_vs_sigma}
Even torkance $t^{\rm e}_{ij}$ 
in the Co(3)/Cu(9) magnetic bilayer
vs.\ lifetime broadening $\Gamma$
when $\theta=\phi=0$.
}
\end{figure}

For comparison, we show in Fig.~\ref{eventorque_cocubilayer_vs_sigma}
the total even torkance (see Eq.~\eqref{eq_sot_total}) 
in the Co(3)/Cu(9) bilayer as a function of
lifetime broadening $\Gamma$.
At large broadening $\Gamma=$100meV 
$t^{\rm top,e}_{xy}$ in the Co(3)/Cu(9)/Co(3) trilayer
is larger than
$t^{\rm e}_{xy}$ in the Co(3)/Cu(9) bilayer
by 10\% while at small
broadening $\Gamma=$1meV
$t^{\rm top,e}_{xy}$ in the Co(3)/Cu(9)/Co(3) trilayer
is larger than
$t^{\rm e}_{xy}$ in the Co(3)/Cu(9) bilayer
by 24\%. This trend can be explained again
by assuming that the effect of the bottom FM
on the top FM is mediated by spin-currents and
that this contribution reduces when the disorder increases. 
Similarly, 
$t^{\rm top,e}_{yx}$ in the Co(3)/Cu(9)/Co(3) trilayer
is larger than
$t^{\rm e}_{yx}$ in the Co(3)/Cu(9) bilayer
by 66\% at large broadening $\Gamma=$100meV, 
while at small broadening $\Gamma=$1meV
$t^{\rm top,e}_{yx}$ in the Co(3)/Cu(9)/Co(3) trilayer
is larger than
$t^{\rm e}_{yx}$ in the Co(3)/Cu(9) bilayer
by 113\%.

\begin{figure}
\includegraphics[width=\linewidth]{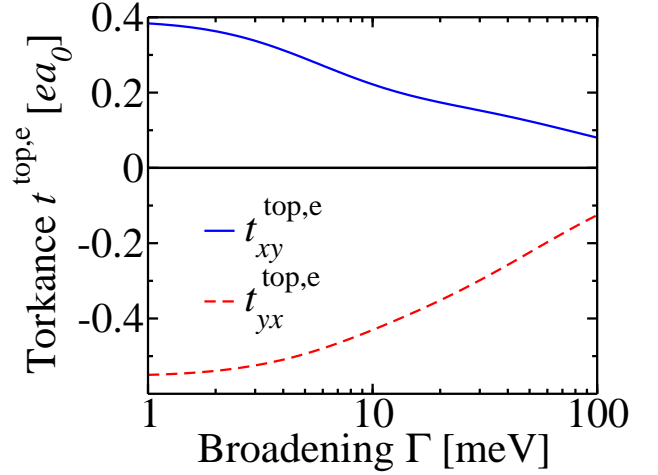}
\caption{\label{eventorque_vs_sigma_nosoicu}
Even torkance $t^{\rm top,e}_{ij}$
on the top magnetic layer 
in the Co(3)/Cu(9)/Co(3) magnetic trilayer
vs.\ lifetime broadening $\Gamma$
when $\theta=\phi=0$ and when SOI on the
Cu atoms is switched off.
}
\end{figure}

Since the SOI in Cu is small, one may expect 
that it does not contribute to the even torque. In order
to verify this by calculations we switch off SOI on all Cu atoms. 
The resulting even torkance on the top magnetic layer
is shown in Fig.~\ref{eventorque_vs_sigma_nosoicu} 
as a function of broadening $\Gamma$ when magnetization
points in $z$ direction ($\theta=0$).
In comparison with Fig.~\ref{eventorque_cocucotrilayer_vs_sigma}
the even torkances are quite similar. Therefore, while switching off SOI on Cu
obviously has an effect on the torkances it is certainly not the dominant
source of $t_{ij}^{\rm top,e}$.
This is a major difference to Co/Pt bilayers, where SOI in Pt provides a strong SHE.

\begin{figure}
\includegraphics[width=\linewidth]{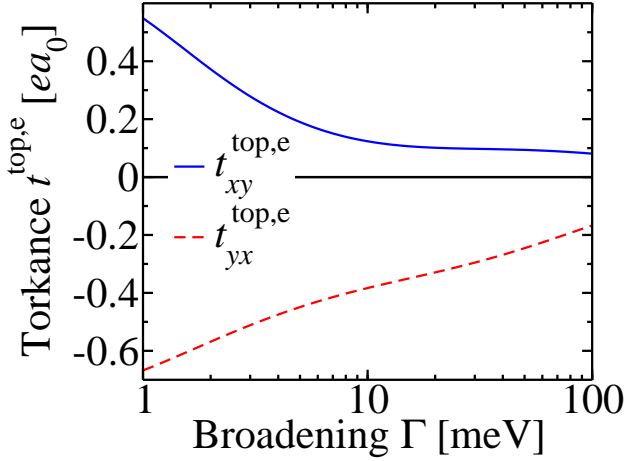}
\caption{\label{eventorque_3Co3Cu3Co}
Even torkance $t^{\rm top,e}_{ij}$ 
in the Co(3)/Cu(3)/Co(3) magnetic trilayer
vs.\ lifetime broadening $\Gamma$
when $\theta=\phi=0$.
}
\end{figure}

\begin{figure}
\includegraphics[width=\linewidth]{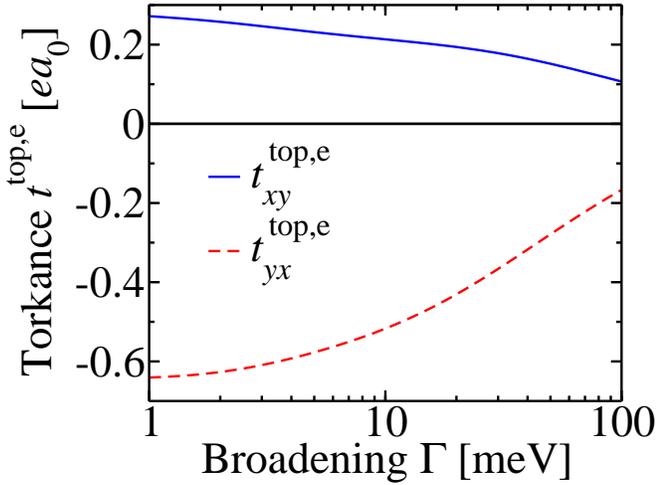}
\caption{\label{eventorque_3Co6Cu3Co}
Even torkance $t^{\rm top,e}_{ij}$ 
in the Co(3)/Cu(6)/Co(3) magnetic trilayer
vs.\ lifetime broadening $\Gamma$
when $\theta=\phi=0$.
}
\end{figure}

In order to investigate the dependence on the thickness of the
Cu layer, we show in Fig.~\ref{eventorque_3Co3Cu3Co} the
even torkance $t^{\rm top,e}_{ij}$ of the Co(3)/Cu(3)/Co(3) trilayer,
in which the spacer consists of only 
3 atomic layers of Cu, and in Fig.~\ref{eventorque_3Co6Cu3Co}
we show 
the even torkance $t^{\rm top,e}_{ij}$  
of Co(3)/Cu(6)/Co(3), which contains
6 layers of Cu in the spacer layer.
At large broadening of $\Gamma=100$~meV the three systems
Co(3)/Cu(3)/Co(3), Co(3)/Cu(6)/Co(3), and Co(3)/Cu(9)/Co(3)
exhibit very similar even torkances, but at smaller broadenings 
the torkances vary differently with $\Gamma$. That the torkance
becomes independent of the layer thicknesses when $\Gamma$ 
is sufficiently large has been found also for Co/Pt bilayers~\cite{ibcsoit}.
The component $t^{\rm top,e}_{yx}$ is 
enhanced in comparison to $t^{\rm top,e}_{xy}$ in both 
Co(3)/Cu(3)/Co(3) and Co(3)/Cu(6)/Co(3) at small broadening $\Gamma$, 
similar to Co(3)/Cu(9)/Co(3) discussed above.

\begin{figure}
\includegraphics[width=\linewidth]{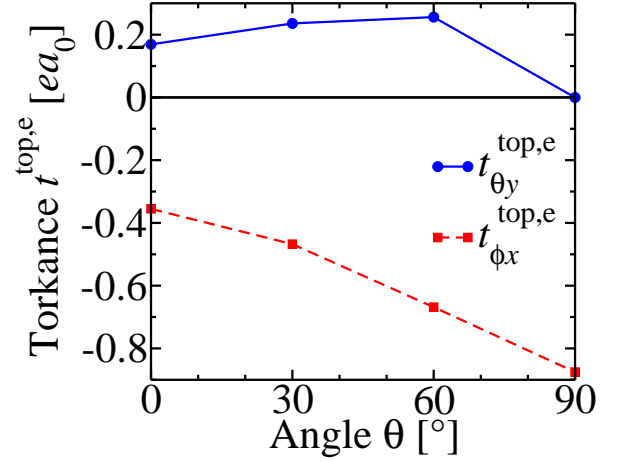}
\caption{\label{eventorque_vs_angle}
Even torkance $t^{\rm top,e}_{\theta y}$ (circles)
and  even torkance $t^{\rm top,e}_{\phi x}$ (squares)
in the Co(3)/Cu(9)/Co(3) magnetic trilayer
vs.\ angle $\theta$
when $\phi=0$ and $\Gamma=25$meV. 
Lines serve as guide to the eye.
}
\end{figure}

Next, we discuss how the even SOT depends on the direction of $\magdir^{\rm top}$.
Since the torque is always perpendicular
to the magnetization, it is convenient to discuss it using the
unit vectors of the spherical
coordinate system
\bege
\hat{\vn{e}}_{\theta}
=(\cos\theta\cos\phi,\cos\theta\sin\phi,-\sin\theta)^{\rm T}
\ee
and
\bege
\hat{\vn{e}}_{\phi}=
(-\sin\phi,\cos\phi,0)^{\rm T}.
\ee
We define the torkances $t^{\rm top}_{\theta i}$ 
and $t^{\rm top}_{\phi i}$ through
\bege\label{eq_theta_component}
\vn{T}^{\rm top}\cdot \hat{\vn{e}}_{\theta}
=\sum_{i}t^{\rm top}_{\theta i}E_{i}
\ee
and
\bege\label{eq_phi_component}
\vn{T}^{\rm top}\cdot \hat{\vn{e}}_{\phi}
=\sum_{i}t^{\rm top}_{\phi i}E_{i}.
\ee
In Fig.~\ref{eventorque_vs_angle} we show the even torkance 
in the Co(3)/Cu(9)/Co(3) trilayer as
a function of the angle $\theta$ when $\phi=0$ 
and $\Gamma=25$meV.
In agreement with the symmetry
analysis in Table~\ref{tab_symmetry2} the even SOT is 
zero for $\theta=90^{\circ}$ when the electric current is applied 
in $y$ direction. However, when the electric field is applied in $x$ direction,
the even torkance increases strongly 
from $t^{\rm e}_{\phi x}=-0.36 e a_{0}$ (at $\theta=0^{\circ}$)
to $t^{\rm e}_{\phi x}=-0.88 e a_{0}$ (at $\theta=90^{\circ}$).
A similarly strong increase of the even torkance with the magnetization
angle has been found in experiments on MgO/CoFeB/Ta~\cite{symmetry_spin_orbit_torques}.

\begin{figure}
\includegraphics[width=\linewidth]{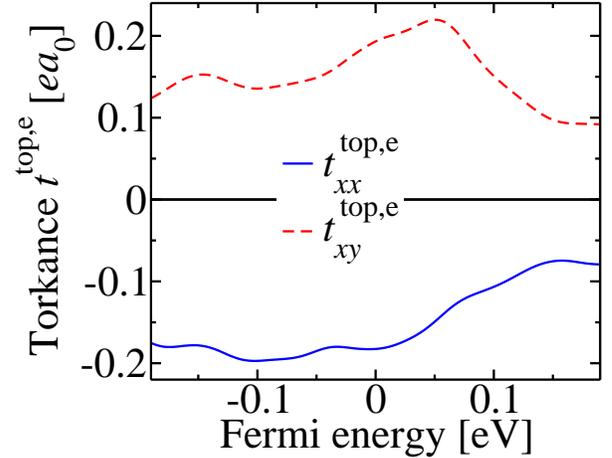}
\caption{\label{eventorque_theta90_phi90}
Even torkance $t^{\rm top,e}_{ij}$ 
in the Co(3)/Cu(9)/Co(3) magnetic trilayer
vs.\ Fermi energy $\mathcal{E}_{\rm F}$ when
$\theta=\phi=90^{\circ}$ and $\Gamma=25$meV.
}
\end{figure}

In Fig.~\ref{eventorque_theta90_phi90} we show the even torkance as
a function of Fermi energy $\mathcal{E}_{\rm F}$ 
when $\theta=\phi=90^{\circ}$.
In agreement with the symmetry
analysis in Table~\ref{tab_symmetry2} the even SOT 
points in $x$ direction in this case and it can be generated
by an electric field in $x$ direction, but also by an electric field in $y$ direction.
As discussed at the end of section Sec.~\ref{sec_symmetry}, $t^{\rm top,e}_{xx}$ 
can be explained by a spin current with spin-polarization in $z$ direction,
which flows from the bottom magnet into the top magnet, and which precesses
around the top FM magnetization, thereby generating a torque
in the direction $\hat{\vn{M}}^{\rm top}\times\hat{\vn{e}}_{z}=\hat{\vn{e}}_{y}\times\hat{\vn{e}}_{z}=\hat{\vn{e}}_{x}$.

\begin{figure}
\includegraphics[width=\linewidth]{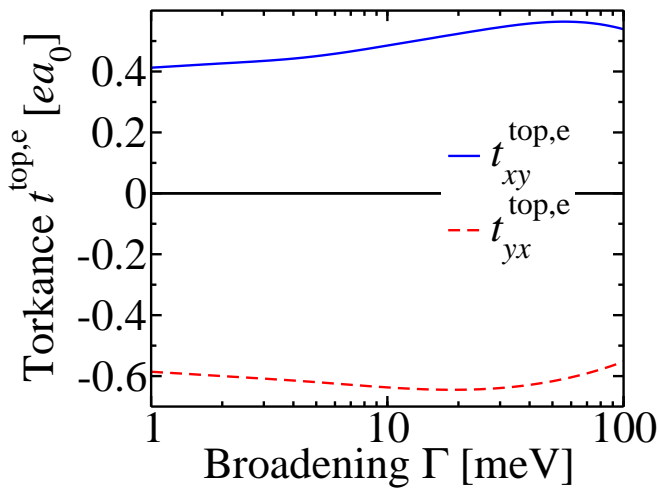}
\caption{\label{eventorque_co3pt13co3}
Even torkance $t^{\rm top,e}_{ij}$ 
in the Co(3)/Pt(13)/Co(3) magnetic trilayer
vs.\ broadening $\Gamma$ when
$\theta=\phi=0^{\circ}$.
}
\end{figure}

In order to compare the SOTs in Co/Cu/Co trilayers, which 
contain only 3$d$ transition metals, to Co/Pt/Co trilayers,
in which Pt provides a large SHE,  
we show in Fig.~\ref{eventorque_co3pt13co3} the
even torkance in Co/Pt/Co 
as a function of broadening $\Gamma$
when $\theta=\phi=0$. Similar to the Co/Cu/Co trilayers, $t^{\rm top,e}_{yx}$
is enhanced when compared to $t^{\rm top,e}_{xy}$ at
small broadening $\Gamma$ due to the presence of the bottom magnet.
The suppression of the SOT with increasing $\Gamma$ is considerably
smaller in the  Co/Pt/Co trilayers than in the Co/Cu/Co trilayers, consistent with
our interpretation that the SOT in  Co/Cu/Co trilayers arises from different
mechanisms.
For small broadening $\Gamma$ the even torkances in Co/Cu/Co and Co/Pt/Co
are of the same order of magnitude.

At first glance it is suprising that the even torkances in Co/Cu/Co trilayers are comparable 
in magnitude to those in Co/Pt bilayers~\cite{ibcsoit,invsot}
and to those in the Co/Pt/Co trilayers considered here, 
despite the absence of any 5$d$ or 4$d$ transition metals with
strong SOI in the former.
However, it is well-known that several SOI-generated effects
can be large even in systems that do not contain any 5$d$ or 4$d$ transition metals.
For example, the intrinsic AHE
of bcc Fe~\cite{fipri_ahe_fe} is not much smaller than
the one of $L$1$_{0}$ FePt~\cite{ahe_FePd_FePt}, despite the presence
of a 5$d$ element in the latter.
Also in the Heusler compound Co$_2$MnAl the AHE is large even though
it is composed only of 3$d$ transition metals~\cite{ahe_heuslers_felser}.
Similarly, the MAE in Fe/MgO~\cite{fipri_mae_FeMgO}
is found to be large, despite the absence of 4$d$ and 5$d$ elements
with strong SOI.

According to Eq.~\eqref{eq_torque_and_torkance} 
the torkance specifies the torque per electric
field. 
However, in order to judge whether the SOT
in a given system is sufficiently large to make it
attractive for applications it is important
to know the torque per electric current. 
The torque per electric current is
obtained from the torkance by multiplication with 
the electrical resistivity, i.e., $T_{y}/J_{x}=t_{yx}\rho_{xx}$ for example.
Since the resistivity of bulk Cu is much lower than the one of bulk Pt we estimate
that the even torque per current in Co(3)/Cu(9) bilayers 
and in Co(3)/Cu(9)/Co(3) trilayers
is smaller by roughly one order of magnitude when compared to
Co/Pt or Co/Pt/Co. 

This amplification by the longitudinal resistivity $\rho_{xx}$ is
also known from the anomalous Hall effect and from the spin
Hall effect, where materials with large Hall angles are most
attractive for applications. The Hall angle is the product of the
Hall conductivity and the longitudinal resistivity $\rho_{xx}$.
The antiferromagnets MnIr, MnPt and MnPd have spin Hall angles comparable
to the one of Pt despite smaller spin Hall conductivities, 
because they have larger resistivities than Pt~\cite{she_in_afms}. 
Also the large spin Hall angles
of $\beta$-W and $\beta$-Ta profit from the high longitudinal 
resistivity $\rho_{xx}$ of 
the $\beta$-phase~\cite{stt_devices_giant_she_tungsten,spin_torque_giant_she_tantalum}. 
In topological insulators one can even achieve infinitely large spin Hall angles,
at least theoretically, and thereby generate large 
SOTs~\cite{spin_transfer_torque_generated_by_topological_insulator}.
Therefore, our finding that the even torkances
in Co(3)/Cu(9)/Co(3) and Co(3)/Cu(9) are comparable in size to
those in Co/Pt and Co/Pt/Co signifies that one may expect sufficiently large
torque per current ratios even when only 3$d$ transition metals are
involved, provided that the longitudinal resistivity $\rho_{xx}$ 
is large enough. While we chose
for the spacer layer in this study Cu, which is known to be a good conductor in bulk, it
can easily be replaced for example by Ti. 
Alternatively, one may increase
the resistivity of Cu by alloying and thereby increase the torque per current
ratio.
In this sense Co(3)/Cu(9)/Co(3) and Co(3)/Cu(9) can be considered
as models of trilayers and bilayers that are composed of 3$d$ transition
metals only. 
In fact, SOT-switching has recently been demonstated 
in FM/Ti/CoFeB trilayers~\cite{sots_interface_generated_spin_currents_baek},
which shows that sufficiently large torque per current ratios can indeed be realized even
when only 3$d$ transition metals are used.

\subsection{Odd SOT}
\begin{figure}
\includegraphics[width=\linewidth]{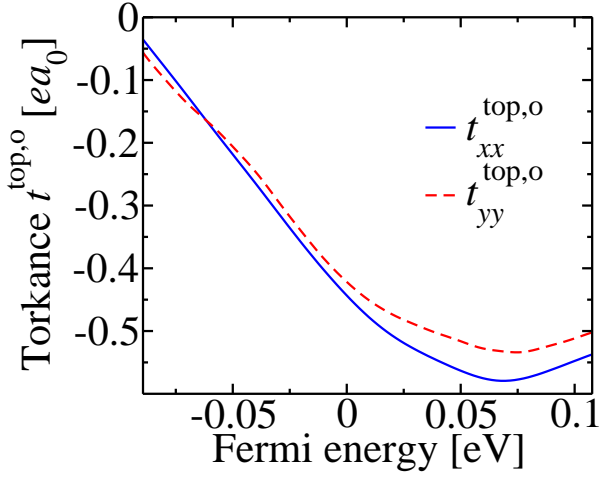}
\caption{\label{oddtorque_toplayer_vs_fermi_cocuco}
Odd torkance $t^{\rm top,o}_{ij}$ on 
the top magnet in the Co(3)/Cu(9)/Co(3) trilayer
vs.\ Fermi energy $\mathcal{E}_{\rm F}$ when 
$\theta=\phi=0$ and $\Gamma=25$meV.
}
\end{figure}

In Fig.~\ref{oddtorque_toplayer_vs_fermi_cocuco} we show the
odd torkance of the top FM (see Eq.~\eqref{eq_sot_toplayer}) 
in the Co(3)/Cu(9)/Co(3) trilayer
as a function of the Fermi energy $\mathcal{E}_{\rm F}$
for a lifetime broadening of $\Gamma=25$meV when the magnetization
of the top FM points in $z$ direction.
In agreement with the symmetry analysis in Table~\ref{tab_symmetry2}
the nonzero components of the odd torkance are $t_{xx}^{\rm top,o}$ and $t_{yy}^{\rm top,o}$.
There is a small difference between $t_{xx}^{\rm top,o}$ and $t_{yy}^{\rm top,o}$,
i.e., the odd torkance is slightly anisotropic due to the presence of the
bottom magnetic layer with magnetization in $x$ direction, which lowers the
symmetry. However, this anisotropy is much smaller than in the case of the
even torque discussed above in Fig.~\ref{cocuco_eventorque_vs_fermi}.

\begin{figure}
\includegraphics[width=\linewidth]{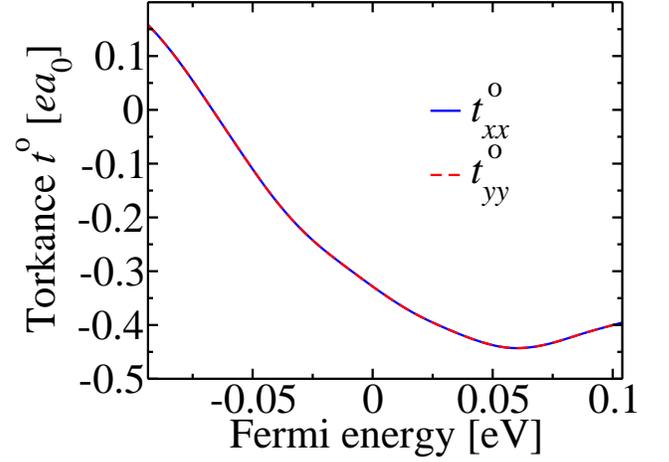}
\caption{\label{oddtorque_vs_fermi_cocu}
Odd torkance $t^{\rm o}_{ij}$ in
the Co(3)/Cu(9) bilayer 
vs.\ Fermi energy $\mathcal{E}_{\rm F}$
when $\theta=\phi=0$ and $\Gamma=25$meV.
}
\end{figure}

For comparison, we show in 
Fig.~\ref{oddtorque_vs_fermi_cocu} the
total odd torkance (see Eq.~\eqref{eq_sot_total}) in the Co(3)/Cu(9) bilayer
as a function of the Fermi energy $\mathcal{E}_{\rm F}$
for a lifetime broadening of $\Gamma=25$meV when the magnetization
points in $z$ direction.
In this case, the odd torkance in the Co(3)/Cu(9) bilayer is 
isotropic, i.e., $t_{xx}^{\rm o}=t_{yy}^{\rm o}$.
The odd torkance in Co(3)/Cu(9) is
very similar to the one
in the Co(3)/Cu(9)/Co(3) trilayer.
Thus, the influence of the bottom magnetic layer on the odd torkance
is much smaller than its influence on the even torkance. 
This shows that the
odd SOT is dominantly local, i.e., it arises from the SOI in the top FM
and from the interfacial SOI at the top Co/Cu interface (below we will
see that for $\theta=\phi=90^{\circ}$ an additional nonlocal contribution
to the odd torque can be observed). 
In contrast, the even SOT 
in Co(3)/Cu(9)/Co(3) contains an important
nonlocal contribution mediated by spin-currents that couple the
bottom FM and the top FM.  
In the pure Co(3)/Cu(9) bilayer ($\mathcal{E}_{\rm F}=0$) the odd torkance
amounts to $t_{xx}^{\rm o}=0.33ea_{0}$. Interestingly, this value is larger
by roughly a factor of four than the odd torkance for a monolayer of Co on Cu(111)
obtained from a similar approach 
with $\Gamma=$25meV (see Fig.~8(b) in Ref.~\cite{PhysRevB.93.224420}).
We attribute this difference to the 
different interfaces (in the present work
the interface normal direction corresponds to the (001) direction of Co and Cu,
while in Ref.~\cite{PhysRevB.93.224420} the interface normal direction
corresponds to the (111) direction of Co and Cu)
and to the
different thicknesses of the Co layer (1 layer of Co in 
Ref.~\cite{PhysRevB.93.224420} compared
to three layers of Co in the present study). 

\begin{figure}
\includegraphics[width=\linewidth]{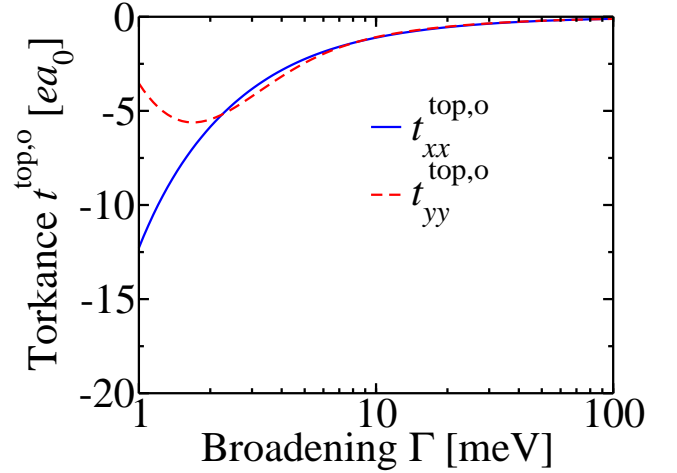}
\caption{\label{oddtorque_toplayer_vs_sigma_cocuco}
Odd torkance $t^{\rm top,o}_{ij}$ on 
the top magnet in the Co(3)/Cu(9)/Co(3) trilayer
vs.\ lifetime broadening $\Gamma$ 
when $\theta=\phi=0$.
}
\end{figure}

In Fig.~\ref{oddtorque_toplayer_vs_sigma_cocuco} we show the
odd torkance of the top FM (see Eq.~\eqref{eq_sot_toplayer}) 
in the Co(3)/Cu(9)/Co(3) trilayer
as a function of the lifetime broadening $\Gamma$.
At large broadening of $\Gamma=$100meV 
the component $t_{xx}^{\rm top,o}$ is smaller
than $t_{yy}^{\rm top,o}$ by 3\%.
At medium broadening of $\Gamma=$10meV  
the component $t_{xx}^{\rm top,o}$ is larger
than $t_{yy}^{\rm top,o}$ by 2.7\%.
At small broadening of $\Gamma=$1meV  
the component $t_{xx}^{\rm top,o}$ is larger
than $t_{yy}^{\rm o}$ by a factor of 3.5.
Thus, like in the case of the even SOT,
we find that the anisotropy of the torkance
with respect to the in-plane rotation of the
applied electric field reduces with
increasing lifetime broadening $\Gamma$.
However, the reduction of this anisotropy
with increasing $\Gamma$ is much more abrupt
than in the case of the even SOT and cannot
be explained by the reduction of the
spin-diffusion length. We attribute the strong
anisotropy of the odd SOT at small $\Gamma$
to the anisotropy of the Fermi surface,
which is induced by the bottom FM.
For medium and large values of $\Gamma$
the odd SOT is less sensitive to this Fermi surface
anisotropy than for small $\Gamma$.

\begin{figure}
\includegraphics[width=\linewidth]{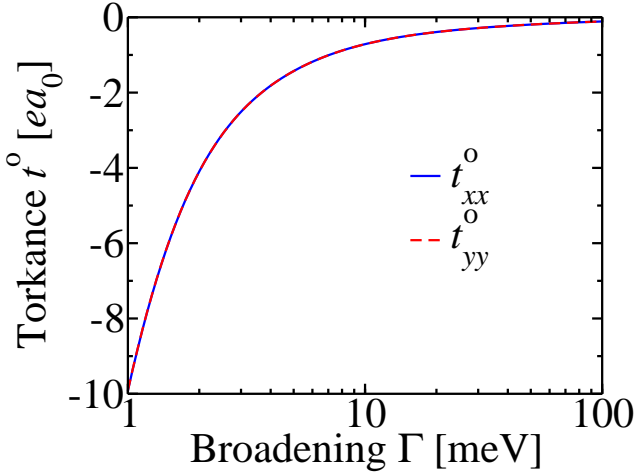}
\caption{\label{oddtorque_vs_sigma_cocu}
Odd torkance $t^{\rm o}_{ij}$ in
the Co(3)/Cu(9) bilayer
vs.\ lifetime broadening $\Gamma$
when $\theta=\phi=0$.
}
\end{figure}

For comparison, we show in
Fig.~\ref{oddtorque_vs_sigma_cocu} the
total odd torkance (see Eq.~\eqref{eq_sot_total}) in the Co(3)/Cu(9) bilayer
as a function of the lifetime broadening $\Gamma$.
At large broadening $\Gamma=$100meV
$t^{\rm o}_{xx}$ in the Co(3)/Cu(9) bilayer
is larger than
$t^{\rm top,o}_{xx}$ in the Co(3)/Cu(9)/Co(3) trilayer
by 1.7\% while at small
broadening $\Gamma=$1meV
$t^{\rm top,o}_{xx}$ in the Co(3)/Cu(9)/Co(3) trilayer
is larger than
$t^{\rm o}_{xx}$ in the Co(3)/Cu(9) bilayer
by 24\%. 
In contrast,
$t^{\rm top,o}_{yy}$ in the Co(3)/Cu(9)/Co(3) trilayer
is larger than
$t^{\rm o}_{yy}$ in the Co(3)/Cu(9) bilayer
by 1.5\% at large broadening $\Gamma=$100meV,
while at small broadening $\Gamma=$1meV
$t^{\rm o}_{yy}$ in the Co(3)/Cu(9) bilayer
is larger than
$t^{\rm top,o}_{yy}$ in the Co(3)/Cu(9)/Co(3) trilayer
by a factor of 2.8. Thus, the odd torques in the
Co(3)/Cu(9)/Co(3) trilayer are almost identical to those
in the Co(3)/Cu(9) bilayer when the broadening $\Gamma$ is large,
while for small broadening in particular the $yy$ component of
the torkance tensor differs significantly between the bilayer
and the trilayer.

\begin{figure}
\includegraphics[width=\linewidth]{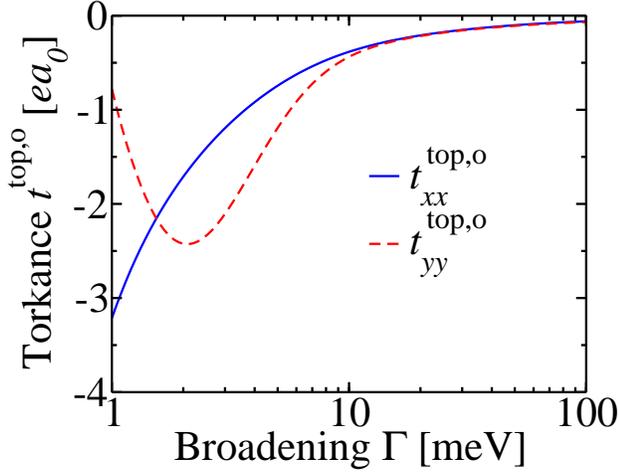}
\caption{\label{oddtorque_vs_gamma_nosoicu}
Odd torkance $t^{\rm top, o}_{ij}$ 
in the Co(3)/Cu(9)/Co(3) magnetic trilayer
vs.\ lifetime broadening $\Gamma$ when
$\theta=\phi=0$ and when SOI is switched off on the
Cu atoms.
}
\end{figure}

In order to discuss the contribution of SOI in Cu to the
odd torque we show in Fig.~\ref{oddtorque_vs_gamma_nosoicu}
the torkance $t^{\rm top, o}_{ij}$ as a function of lifetime
broadening $\Gamma$ when SOI is switched off on the Cu atoms.
Compared to Fig.~\ref{oddtorque_toplayer_vs_sigma_cocuco}
the torkance is strongly decreased when SOI is switched off on the Cu atoms.
This is a remarkable difference to the even torques, for which we discussed
above that switching off SOI on Cu has a small influence.

\begin{figure}
\includegraphics[width=\linewidth]{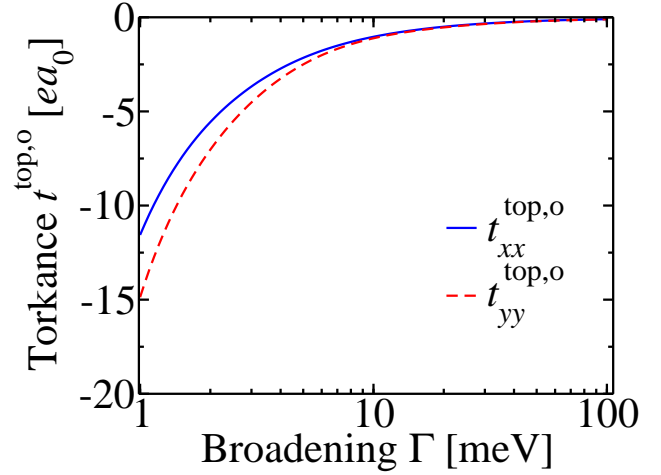}
\caption{\label{fig_oddtorque_3Co3Cu3Co_vs_sigma}
Odd torkance $t^{\rm top, o}_{ij}$ 
in the Co(3)/Cu(3)/Co(3) magnetic trilayer
vs.\ lifetime broadening $\Gamma$ when
$\theta=\phi=0$.
}
\end{figure}

\begin{figure}
\includegraphics[width=\linewidth]{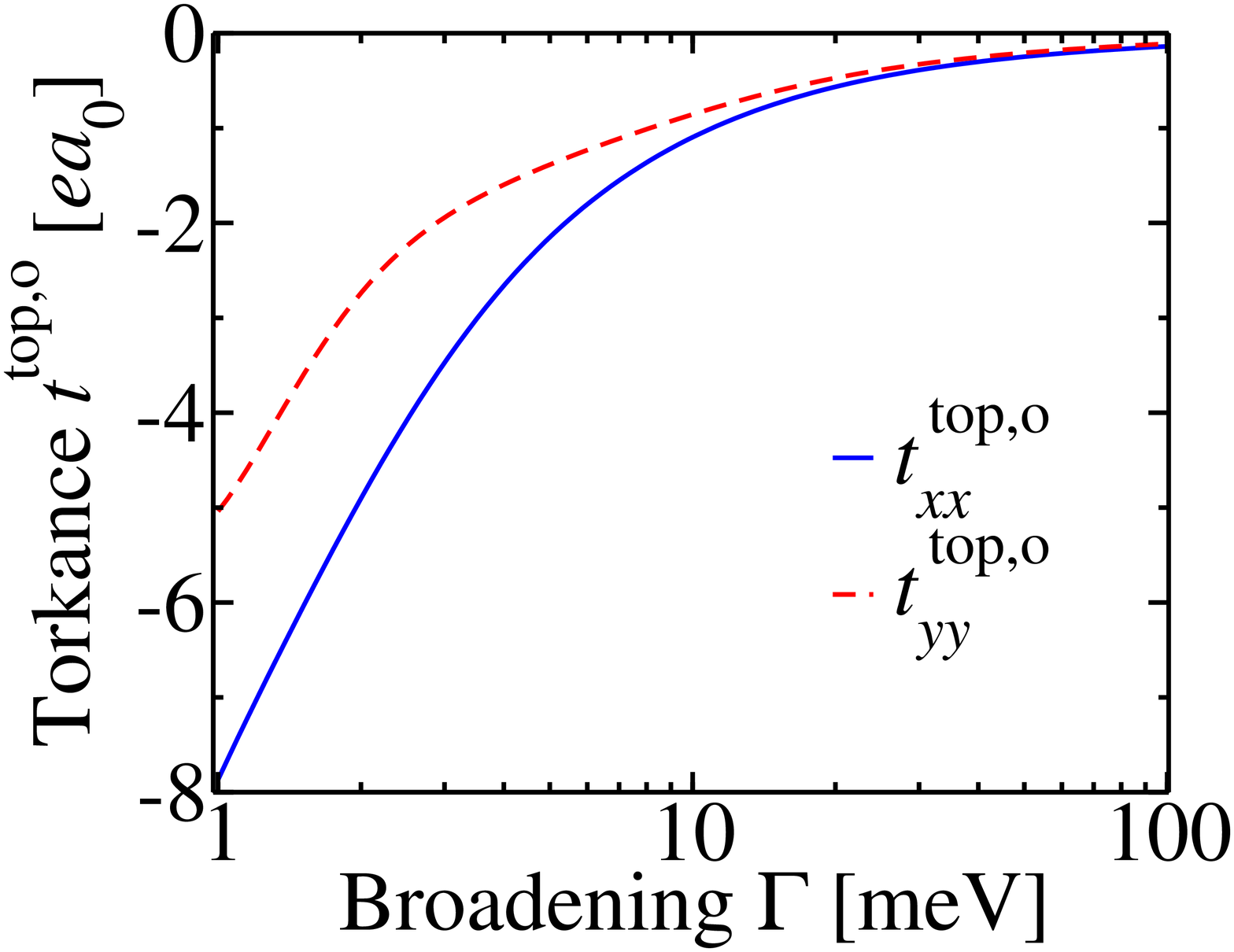}
\caption{\label{fig_oddtorque_3Co6Cu3Co_vs_sigma}
Odd torkance $t^{\rm top, o}_{ij}$ 
in the Co(3)/Cu(6)/Co(3) magnetic trilayer
vs.\ lifetime broadening $\Gamma$ when
$\theta=\phi=0$.
}
\end{figure}

In order to investigate the dependence of the odd torque on
the thickness of the Cu spacer layer, we show the odd torkances
of Co(3)/Cu(3)/Co(3) and Co(3)/Cu(6)/Co(3) in 
Fig.~\ref{fig_oddtorque_3Co3Cu3Co_vs_sigma}
and in Fig.~\ref{fig_oddtorque_3Co6Cu3Co_vs_sigma},
respectively.
Like in the case of the even torkance discussed in the
previous subsection, the
odd torkances of Co(3)/Cu(3)/Co(3), Co(3)/Cu(6)/Co(3), and
Co(3)/Cu(9)/Co(3) are similar for sufficiently large $\Gamma$
and differ substantially for small $\Gamma$.

\begin{figure}
\includegraphics[width=\linewidth]{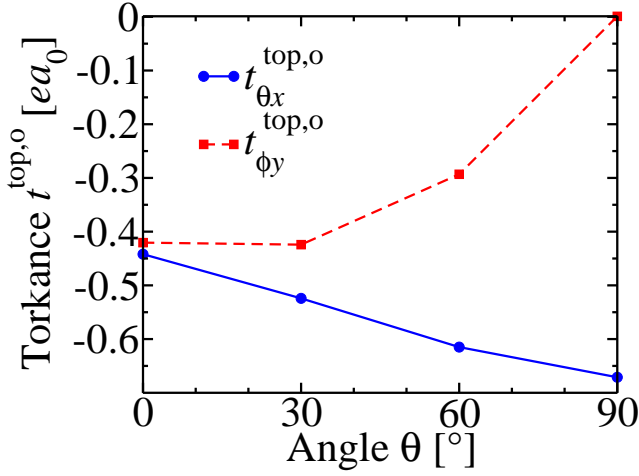}
\caption{\label{oddtorque_vs_angle}
Odd torkance $t^{\rm top, o}_{\theta x}$ (circles)
and odd torkance $t^{\rm top, o}_{\phi y}$ (squares) 
in the Co(3)/Cu(9)/Co(3) magnetic trilayer
vs.\ angle $\theta$ when $\phi=0$ and $\Gamma=25$meV. 
Lines serve as guide to the eye.
}
\end{figure}

In Fig.~\ref{oddtorque_vs_angle} we show the odd torkance 
in the Co(3)/Cu(9)/Co(3) trilayer as
a function of the angle $\theta$ when $\phi=0$ and $\Gamma=25$meV. 
We plot the $\theta$ and $\phi$ components of the torkance defined in
Eq.~\eqref{eq_theta_component} and
in Eq.~\eqref{eq_phi_component},
respectively.
The torkance $t^{\rm top,o}_{\phi y}$ vanishes at $\theta=90^{\circ}$ in
agreement with the symmetry analysis in Table~\ref{tab_symmetry2}.
However, $t^{\rm top,o}_{\theta x}$ increases strongly
from $t^{\rm top,o}_{\theta x}=-0.44 e a_{0}$ ($\theta=0^{\circ}$)
to $t^{\rm top,o}_{\theta x}=-0.67 e a_{0}$ ($\theta=90^{\circ}$).
As discussed above, a similar increase occurs for the even 
torkance $t^{\rm top,e}_{\phi x}$ shown in Fig.~\ref{eventorque_vs_angle}.
One contribution to the increase of $t^{\rm top,o}_{\theta x}$ with 
the angle $\theta$ may come from the $z$-polarized spin current present
in trilayers, because it is expected to provide a term proportional to $\sin(\theta)$.
However, experimentally a strong angular dependence of the odd torque is
also found in bilayers such as AlO$_{x}$/Co/Pt 
and MgO/CoFeB/Ta~\cite{symmetry_spin_orbit_torques}.
\begin{figure}
\includegraphics[width=\linewidth]{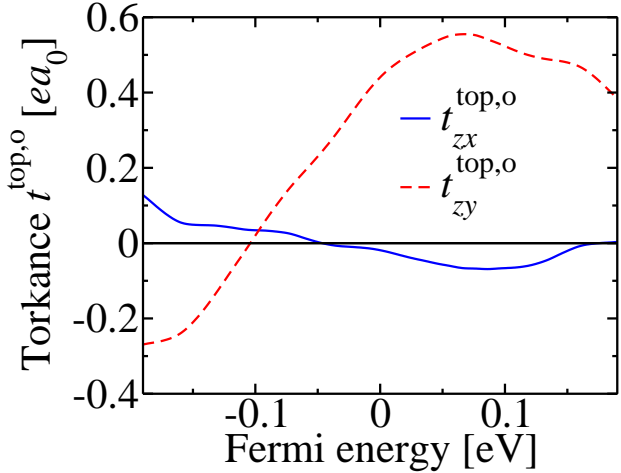}
\caption{\label{oddtorque_theta90_phi90}
Odd torkance $t^{\rm top,o}_{ij}$ 
in the Co(3)/Cu(9)/Co(3) magnetic trilayer
vs.\ Fermi energy $\mathcal{E}_{\rm F}$ when
$\theta=\phi=90^{\circ}$ and $\Gamma=25$meV.
}
\end{figure}

In Fig.~\ref{oddtorque_theta90_phi90} we show the odd torkance as
a function of Fermi energy $\mathcal{E}_{\rm F}$ 
when $\theta=\phi=90^{\circ}$.
In agreement with the symmetry
analysis in Table~\ref{tab_symmetry2} the odd SOT 
points in $z$ direction in this case, and it can be 
generated by an electric field in $x$ direction, but also by an
electric field in $y$ direction. 
As discussed at the end of Sec.~\ref{sec_symmetry},
we attribute $t^{\rm top,o}_{zx}$ to a spin current
with spin-polarization along $z$ direction flowing from the
bottom Co layer to the top Co layer and assume that this
spin current is generated at the bottom FM interface through
spin-orbit precession.
$t^{\rm top,o}_{zx}$ is small at $\mathcal{E}_{\rm F}=0$ and 
therefore also the $z$-polarized spin current is small. The $z$-polarized
spin current therefore provides only a small contribution to the
$\theta$-dependence of  $t^{\rm top,o}_{\theta x}$ shown in 
Fig.~\ref{oddtorque_vs_angle}.

\begin{figure}
\includegraphics[width=\linewidth]{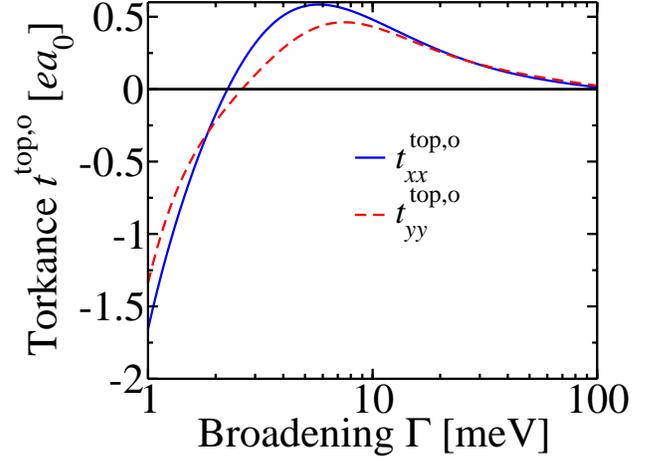}
\caption{\label{coptco_oddtorque_vs_sigma}
Odd torkance $t^{\rm top,o}_{ij}$ 
in the Co(3)/Pt(13)/Co(3) magnetic trilayer
vs.\ broadening $\Gamma$ when
$\theta=\phi=0^{\circ}$.
}
\end{figure}

Finally, we discuss the odd torque in the Co(3)/Pt(13)/Co(3) trilayer.
In Fig.~\ref{coptco_oddtorque_vs_sigma} we show the odd torkance
as a function of the lifetime broadening $\Gamma$ when $\theta=\phi=0^{\circ}$.
Similar to the Co/Cu/Co trilayers, the difference between $t^{\rm top,o}_{xx}$
and $t^{\rm top,o}_{yy}$ is pronounced only at small $\Gamma$.
Remarkably, for small and medium values of $\Gamma$ 
the odd torkance in the Co(3)/Pt(13)/Co(3) trilayer is smaller
than the one in the Co/Cu/Co trilayers, showing again that sizable SOTs can
be observed even in trilayers that are composed only of 3$d$ transition metals. 

\section{DMI in C\lowercase{o}/C\lowercase{u}/C\lowercase{o} trilayers}
\label{sec_dmi_in_cocuco}
Since DMI and SOT are related in various 
ways~\cite{mothedmisot,sot_dmi_stiles,common_origin_sot_dmi},
we expect the DMI in Co/Cu/Co trilayers to be anisotropic 
due to the bottom FM 
like
the SOT, the anisotropy of which we discussed above.
We calculate the DMI according to Eq.~\eqref{eq_dmi} for magnetization along $z$ in the
top magnet.
In Table~\ref{table_dmi} we show the layer-resolved DMI coefficients 
for the Co/Cu/Co trilayer
and for the Co/Cu bilayer.
The largest contribution to the DMI arises from Co-1, which is adjacent to vacuum.
The contributions from Co-3, which is adjacent to the copper layer, are significantly
smaller and opposite in sign. The contributions from Co-3 are small, because the SOI
in Cu is small.
As expected, the DMI is anisotropic in Co/Cu/Co, because the magnetization
in the bottom FM points in $x$ direction. 
The DMI free energy density Eq.~\eqref{eq_free_energy_dmi} of a flat right-handed cycloidal 
spin spiral parallel to the $xz$ plane
with wave vector $q$ in $x$ direction is $D_{yx}q$ while the
DMI free energy density of a flat right-handed cycloidal spin spiral
parallel to the $yz$ plane
with wave vector $q$ 
in $y$ direction is $-D_{xy}q$. These two energy densities differ by 7\%.

\begin{threeparttable}[t]
\caption{\label{table_dmi}
Layer-resolved DMI-coefficients for the
Co/Cu/Co trilayer and for the Co/Cu bilayer.
The sums of the contributions from layers
Co-1, Co-2, Co-3 and Cu-1 are listed in
the line $D_{ij}^{\rm top}$ (see Eq.~\eqref{eq_dmi_toplayer}). 
For Co/Cu only $D_{xy}$
is shown, because $D_{yx}=-D_{xy}$ due to 
symmetry in this case. $D_{yx}\ne -D_{xy}$ for Co/Cu/Co,
because the bottom magnet reduces the symmetry.
The DMI coefficients are specified in units of
meV\AA\, per unit cell (uc).
}
\begin{ruledtabular}
\begin{tabular}{cccc}
&\multicolumn{2}{c}{Co/Cu/Co} &\multicolumn{1}{c}{Co/Cu}\\
\cline{2-3}\cline{4-4}
&$D_{xy}$ [meV\AA/uc]  &$D_{yx}$ [meV\AA/uc] 
&$D_{xy}$ [meV\AA/uc]\\
\hline
Co-1 &-3.07 &2.80 &-3.78 \\
Co-2 &-0.878 &1.07  &-0.58\\
Co-3 &1.24 &-0.94  &0.48\\
Cu-1 &-0.017 &0.014  &-0.01\\
$D_{ij}^{\rm top}$ &-2.73 &2.94  &-3.89
\end{tabular}
\end{ruledtabular}
\end{threeparttable}

This anisotropy provides a simple way to tune the DMI. For a
cycloidal spin spiral that propagates in a given fixed direction in the top magnet
one can change the
angle between the spin-spiral wave vector and the bottom FM magnetization
by rotating the latter.
Thereby one can tune the DMI and as a consequence the wavelength of the spin spiral.
While the anisotropy of DMI is only 7\% in Co/Cu/Co according to our calculations,
we expect that larger anisotropies can be realized by optimizing the composition of
the trilayer with the goal of maximizing this effect. 
In systems with anisotropic DMI skyrmions 
are of elliptic shape instead of circular shape.
By rotating the magnetization in the bottom layer one may 
rotate these ellipsoids and thereby excite the skyrmions.

\section{Summary}
\label{sec_summary}
We find that SOT and DMI in Co/Cu bilayers and Co/Cu/Co trilayers are larger
than what one might expect from the small SOI of Cu.
A small torque per current ratio in Co/Cu bilayers is therefore the result of 
the high electrical conductivity of Cu and not the consequence of a 
small SOT torkance.
This is consistent with recent experiments showing SOT-switching in trilayers
that contain only 3$d$ transition-metal elements.
Co/Cu/Co can serve as a model system to study SOTs in
magnetic trilayers that are composed of 3$d$ transition metals only.
The SOT is anisotropic in Co/Cu/Co when the
magnetization of the bottom magnet is in-plane and the magnetization of
the top magnet is out-of-plane, i.e., the SOT depends on the in-plane
direction of the applied electric current.
We find the anisotropy of the even torque 
to be particularly large and attribute it to
spin currents that are generated at the
bottom magnet. 
The even torque is significantly enhanced
when the applied electric field is aligned with the in-plane magnetization of the
bottom magnet. In contrast, when the electric field is applied perpendicularly to the 
in-plane magnetization of the bottom layer, the even torque in Co/Cu/Co trilayers
is found to be similar to the even torque in Co/Cu bilayers.
When the magnetizations of both the bottom and the top magnet are in-plane, but
perpendicular to each other, we observe a nonlocal SOT that is mediated by
a spin current with spin polarization along the out-of-plane direction. 
Additionally, we find that DMI is anisotropic in Co/Cu/Co trilayers, i.e., 
we predict that the width of domain walls and the spin-spiral wave number of spin-spirals 
depend on their in-plane orientation.
Therefore, we expect that the rotation of the magnetization direction of the bottom magnet
in magnetic trilayers similar to Co/Cu/Co provides an easy way to tune the DMI. 

\section*{Acknowledgments}
We gratefully acknowledge computing time on the supercomputers
of J\"ulich Supercomputing Center and RWTH Aachen University
as well as funding by 
Deutsche Forschungsgemeinschaft (MO 1731/5-1).

\bibliography{sotcocucotrila}

\end{document}